\documentclass{sig-alternate-05-2015}

\usepackage{url}
\usepackage{balance}
\usepackage{cite}
\usepackage{amssymb}
\usepackage{amsfonts}
\usepackage{graphicx}
\usepackage{xspace}
\usepackage{listings}
\usepackage{multicol}
\usepackage{multirow}
\usepackage{epsfig}
\usepackage{tikz}
\usepackage{flexisym}
\usepackage{algpseudocode}
\usepackage{algorithm}
\usepackage{formatting/shortcuts}

\definecolor{lightgray}{rgb}{.9,.9,.9}
\definecolor{darkgray}{rgb}{.4,.4,.4}
\definecolor{purple}{rgb}{0.65, 0.12, 0.82}

\lstdefinelanguage{JavaScript}{
  keywords={typeof, new, true, false, catch, function, return, null, catch, switch, var, if, in, while, do, else, case, break},
  keywordstyle=\color{blue}\bfseries,
  ndkeywords={class, export, boolean, throw, implements, import, this},
  ndkeywordstyle=\color{darkgray}\bfseries,
  identifierstyle=\color{black},
  sensitive=false,
  comment=[l]{//},
  morecomment=[s]{/*}{*/},
  commentstyle=\color{purple}\ttfamily,
  stringstyle=\color{red}\ttfamily,
  morestring=[b]',
  morestring=[b]"
}
\makeatletter
\def\@copyrightspace{\relax}
\makeatother

\begin{document}

% Copyright
% \setcopyright{acmcopyright}
%\setcopyright{acmlicensed}
%\setcopyright{rightsretained}
%\setcopyright{usgov}
%\setcopyright{usgovmixed}
%\setcopyright{cagov}
%\setcopyright{cagovmixed}

% DOI
% \doi{10.475/123_4}

% % ISBN
% \isbn{123-4567-24-567/08/06}

% %Conference
% \conferenceinfo{PLDI '13}{June 16--19, 2013, Seattle, WA, USA}

% \acmPrice{\$15.00}

% %
% % --- Author Metadata here ---
% \conferenceinfo{WOODSTOCK}{'97 El Paso, Texas USA}
%\CopyrightYear{2007} % Allows default copyright year (20XX) to be over-ridden - IF NEED BE.
%\crdata{0-12345-67-8/90/01}  % Allows default copyright data (0-89791-88-6/97/05) to be over-ridden - IF NEED BE.
% --- End of Author Metadata ---

\title{Preventing Your Faults from Telling Your Secrets: Defenses against Pigeonhole Attacks}
% \subtitle{[Extended Abstract]
% \titlenote{A full version of this paper is available as
% \textit{Author's Guide to Preparing ACM SIG Proceedings Using
% \LaTeX$2_\epsilon$\ and BibTeX} at
% \texttt{www.acm.org/eaddress.htm}}}

% \numberofauthors{0} %  in this sample file, there are a *total*
% % of EIGHT authors. SIX appear on the 'first-page' (for formatting
% % reasons) and the remaining two appear in the \additionalauthors section.
% %
\author{
Shweta Shinde, Zheng Leong Chua, Viswesh Narayanan, Prateek Saxena\\
       \affaddr{National University of Singapore}\\
       \email{\{shweta24, chuazl, visweshn, prateeks\} @ comp.nus.edu.sg}
}

\maketitle
\begin{abstract}
%
% Abstract here.

New hardware primitives such as Intel SGX secure a
user-level process in presence of an untrusted or compromised OS. Such ``enclaved
execution'' systems are vulnerable to several side-channels, one of
which is the page fault channel. In this paper, we show that
the page fault side-channel has sufficient channel capacity
to extract bits of encryption keys from commodity implementations of
cryptographic routines in \openssl and \libgcrypt  --- leaking $27$\% on average and up to $100$\% 
of the secret bits in many case-studies. To mitigate this, we propose a software-only
defense that masks page fault patterns by determinising the program's memory access
behavior. We show that such a technique can be built into a compiler,
and implement it for a subset of C which is sufficient to handle the cryptographic
routines we study. This defense when implemented generically can have
significant overhead of up to $4000\times$, but with help of developer-assisted
compiler optimizations, the overhead reduces to at most $29.22$\% in our case studies.
Finally, we discuss scope for hardware-assisted defenses, and show one solution
that can reduce overheads to $6.77$\% with support from hardware changes.

% As a second approach we propose contractual
% execution. With a small change to the hardware, this
% defense incurs a performance overhead of $6.77$\% on
% average.

% We analyze popular cryptographic implementation libraries 
% \openssl~\cite{openssl} and \libgcrypt~\cite{libgcrypt} find that 

\end{abstract}




%
%  Use this command to print the description
%
% \printccsdesc

% We no longer use \terms command

% \keywords{ACM proceedings; \LaTeX; text tagging}
\section{Introduction} 
\label{sec:intro}

Operating systems are designed to execute at higher privileges than
applications on commodity systems.  Recently, this model of assuming a
trusted OS has come under question, with the rise of vulnerabilities
targeting privileged software~\cite{venom}. Consequently, new hardware
primitives have emerged to safeguard applications from untrusted
OSes~\cite{flicker, xom, sgx}. One such primitive is Intel SGX's
enclaved execution which supports secure execution of sensitive
applications on an untrusted OS. The SGX hardware guarantees that all the
application memory is secured and the OS cannot access the application content. 
During execution, applications rely on the OS for
memory management, scheduling and other system services. Intel SGX
holds the promise of affording a private virtual address space for a
trusted process that is immune to active probing attacks from the
hostile OS. However, side-channels such as the page-fault channel 
have been recently discovered~\cite{cca-sgx}.  Since the
OS manages the virtual-to-physical page translation tables for the
sensitive application, it can observe all page faults and the faulting
page addresses, which leaks information. These attacks show that
mere  memory access control and encryption is not enough to defend against the OS,
which motivates a systematic study of defense solutions to mitigate
this channel.

In this paper, we first show that the channel capacity of the
page-fault channel is sufficient to extract secret key information in
existing implementations of cryptographic routines (\openssl and
\libgcrypt). Cryptographic routines are vital to reducing the TCB and
enclaved applications are expected to critically rely on them to
establish secure channel with the I/O, filesystem and network
sub-systems~\cite{sgx-apps, haven, vc3}. To perform an attack, the
adversarial OS allocates a minimum number of physical pages to the
sensitive enclave process, such that memory accesses spill out of the
allocated set as much as possible, incurring page faults. We call such
attacks as {\em \attacks}\footnote{ These attacks were
also referred to as controlled-channel attacks in previous work.}
because they force the victim process to
spill outside the allocated physical pages, thereby maximizing the
channel capacity of the observed side-channel. 
%  Note that these
% attacks work even for implementations which are hardened against
% timing and cache side-channel attacks. 
They affect a long line of 
% previous and emerging 
systems such as Intel SGX~\cite{sgx},
InkTag~\cite{inktag}, PodArch~\cite{podarch}, and
OverShadow~\cite{overshadow} which protect application memory.

The page fault channel is much easier for the OS to exploit as
compared to other side-channels.  For example, in case of cache
side-channel, the hardware resources such as size, number of data
entries, eviction algorithm and so on are often fixed. The adversary
has a  limited control on these factors and the observations are
mainly local to small fragments of program logic. On the contrary, in
case of \attacks, adversary is much stronger, adaptive, and controls
the underlying physical resource (the number of physical pages).
Moreover, it can make far more granular clock measurements (both
global and local) by invoking and inducing a fault in the
enclave. To defend applications against this unaddressed threat, we
seek a security property that allows an application to execute on any
input data while being agnostic to changes in the number of pages
allocated. The property assures that the OS cannot glean any sensitive
information by observing page faults.  
We call this property as {\em page-fault obliviousness} (or \property).

In this work, we propose a purely software-based defense against
\attacks to achieve \property. We point out that defenses against time
and cache side-channels do not directly prevent \attacks, and
achieving \property has been an open problem~\cite{cca-sgx}.
% Specifically, we show why simple approaches such as self-paging,
% randomization, and masking timing side-channel briefly conjectured in previous
% papers are insufficient or incur high performance costs~\cite{cca-sgx}. 
Our goal is to guarantee that even if the OS observes the page faults,
it cannot distinguish the enclaved execution under any values for the
secret input variables. Our propose approach is called {\em
  \interpreting}, wherein the enclave application exhibits the same
page fault pattern under all values possible for the secret input
variables. Specifically, we modify the program to pro-actively access
all its input-dependent data and code pages in the same sequence
irrespective of the input. In our empirical case studies, the naive
implementation of \interpreting results in an overhead of about
$705\times$ on an average and maximum $4000\times$!  Therefore, we
propose several optimizations techniques which exploit specific
program structure and makes the overhead statistically insignificant
in $8$ cases, while the worst-case performance is $29.22$\%. All our
defenses are implemented as an extension to the LLVM compiler,
presently handling a subset of C/C++ sufficient to handle the
cryptographic case studies.  Finally, we discuss alternative solutions
for efficient defenses, and suggest a new defense which requires
hardware support, but yields an acceptable worst-case overhead of
$6.67\%$ for our case studies.

% As an alternative, we propose a final  performance-efficient scheme called {\em
% \bucketing}. The insight we use is that rather than a costly proactive
% defense,  we add hardware support so that the enclave can detect \attacks.
% The key challenge in \bucketing is to safely terminate the enclave
% post-detection as the point of termination may leak information. We address
% this challenge and our final scheme results in an
% acceptable overhead of $6.77$\% in benign execution and terminates {\em
% safely} if it detects a malicious OS.

% \paragraph{Results.}
% We analyze two most popular cryptographic implementation libraries ---
% \openssl~\cite{openssl} and \libgcrypt~\cite{libgcrypt}  using our  
% semi-% automated \attack detection framework. Of the $24$ routines we analyzed,  $10$
% of them are vulnerable to \attacks. The information leakage in these
% applications varies from $1.5$\% to $100$\% of bits of 
% information leakage about the secret cryptographic key, thus
% highlighting the impact of \attacks in cryptographic routines. We automatically 
% patch all the applications using our approach and rerun the
% analysis to verify  that \attack is eliminated. 

\paragraph{Contributions.}
We make the following contributions:

\begin{itemize} 
\squish
	\item {\em \Attacks on real cryptographic routines.} We
          demonstrate that the page-fault channel has sufficient
          capacity to extract significant secret information in
          widely-used basic cryptographic implementations (AES, \eddsa, RSA 
          and so on).

	% \item {\em \Property.} We design two
 %          approaches: \interpreting and \bucketing that eliminate 
 %          information leakage via page fault channel. In doing so, we recommend a fix to
 %          Intel SGX design so as to support \property with
 %          minimum overhead.

     \item {\em Defense.} We propose \property and design \interpreting approach that eliminates
          information leakage via page fault channel. 
          %Further, we devise optimization strategies to 
          % In doing so, we recommend a fix to
          % Intel SGX design so as to support \property with
          % minimum overhead.

	\item {\em Optimizations \& System Evaluation.} We apply our
          defense to the vulnerable cryptographic utilities from
          \libgcrypt and \openssl, and devise sound optimizations. In
          our experiments, \interpreting amounts to an average of
          $705\times$\ overhead without optimization, and is reduced
          to an acceptable average and worst case overhead of $29.22\%$ after optimization.  
          % Our \bucketing defense incurs an
          % overhead of $6.77$\% on average, with nearly zero overhead
          % in $2$ cases.
\end{itemize}

% \section{\Attacks}
\section{Pigeonhole Attacks}
\label{sec:problem}
In a non-enclaved environment, the OS is responsible for managing the process
memory. Specifically, when launching the process, the OS creates the page
tables and populates empty entries for virtual addresses specified in the
application binary. When a process begins its execution, none of its virtual
pages are mapped to the physical memory. When the process tries to access a
virtual address, the CPU incurs a page fault. The CPU reports information such
as the faulting address, type of page access,  and so on to the OS on behalf
of the faulting process, and the OS swaps in the content from the disk.
Similarly, the OS deletes the virtual-to-physical mappings when it reclaims
the process physical memory as and when requested or when necessary. Thus, a
benign OS makes sure that the process has sufficient memory for execution,
typically, at least $20$ pages in Linux systems~\cite{utlk}.

\subsection{Benign Enclaved Execution}
\begin{figure}
\centering
\epsfig{file=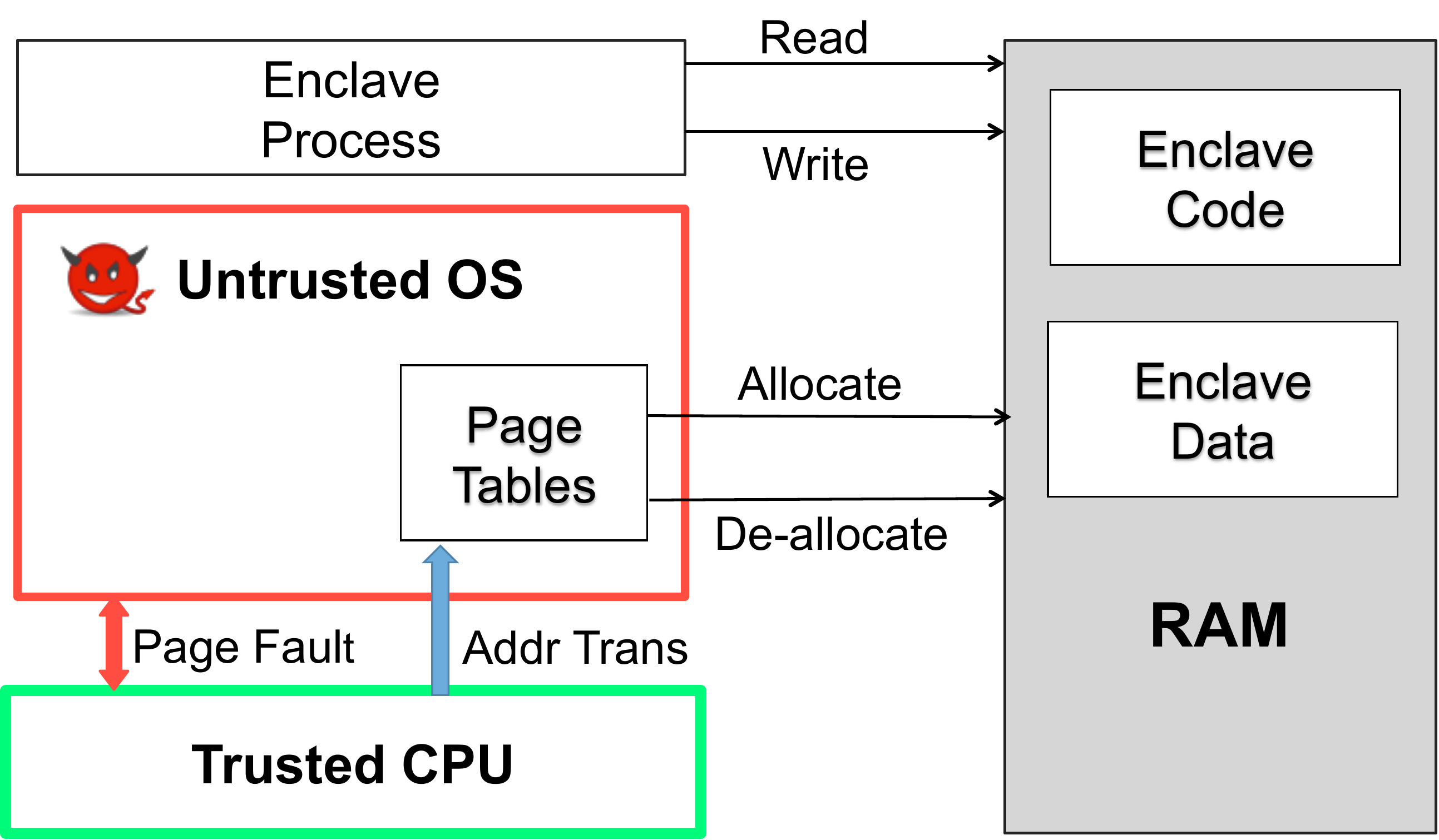, scale=0.29}
\vspace{-0.3cm}
\caption{Problem Setting. Process executing in an enclave on untrusted OS.}
\label{fig:problem}
\vspace{-0.6cm}
\end{figure}

The aim of enclave-like systems is to safeguard all the sensitive process
(called as an enclave) memory during the execution. These systems use memory
encryption and / or memory access controls to preserve the confidentiality of the
sensitive content. The process memory is protected  such that the hardware
allows access in ring-3 only when a legitimate owner process requests to
access its content~\cite{overshadow}. When the OS in ring-0 or any other
process in  ring-3 tries to access the memory, the hardware either encrypts the
content on-demand or denies the access. This guarantees that neither the OS
nor other malicious processes can access the physical memory of an enclave.
In enclaved execution,  the OS memory management functions
are  unchanged. The onus still lies with the OS to decide which process gets
how much physical memory, and which pages should be loaded at which addresses
to maintain the process-OS semantics. The OS controls the page table entries
and is also notified on a page fault.  This CPU design allows the OS to
transparently do its management while the hardware preserves the
confidentiality and integrity of the process memory content.  For example,  if
there are not many concurrent processes executing, the OS may scale up the
memory allocation to a process. Later, the OS may decrease the process memory
when it becomes loaded with memory requests from other processes.  Further,
the CPU reports all the interrupts (such as page fault, general protection
fault) directly to the OS. Figure~\ref{fig:problem} shows the scenario in
\execution, wherein the untrusted OS can use $2$ interfaces:
\texttt{allocate} and \texttt{de-allocate} to directly change the page table for allocating or deallocating process pages respectively. Many systems guarantee secure execution of processes in presence of untrusted
OSes, either at the hardware or software level. Execution of processes in such
isolated environments is referred to as cloaked execution~\cite{overshadow},
\execution~\cite{sgx}, shielded execution~\cite{haven}, and so on depending on
the underlying system.  For simplicity, we refer to all of them as \execution
in this paper, since the underlying mechanism is the same as described
above. See ~\cite{sgx1, sgx2} for SGX-specific details.

\subsection{Pigeonhole Attack via Page Faults}% Side-channel}
In \execution, the OS sees all the virtual addresses where the process
faults~\footnote{In our model, the trusted CPU or hypervisor only reports the
base address of the faulting page while masking the offset within the page
(unlike in InkTag~\cite{inktag}).}. This forms the primary basis of the page
fault side-channel. Each page fault in the \execution  leaks the information
that the process is accessing a specific page at a specific point in execution
time. Since the OS knows the internal structure of the program such as the
layout of the binary, \texttt{mmap}-ed pages, stack, heap, library addresses
and so on, the OS can profile the execution of the program and observe the
page fault pattern. In fact it can invoke and execute the enclave application
for a large number of inputs in offline mode to record the corresponding page fault
patterns.  At runtime, the OS can observe the page  fault pattern for the user
input and map it to its pre-computed database,  thus learning the sensitive
input. The remaining question is, what degree of control does the OS
have on the channel capacity?

An adversarial OS that is actively misusing this side-channel always  aims to
maximize the page faults and extract information for a given input.  On the
upside, applications often follow temporal and spatial locality of reference
and thus do not incur many page faults during execution. Thus, the information
leaked via the benign page faults from the enclave is not significant.
However, note that the adversarial OS controls the process page tables and
decides which virtual pages are to be loaded in the physical memory at a given
point. To perpetrate the \attack, the OS allocates only three pages at most to
the program at a particular moment --- the code page, the source address and
the destination address~\footnote{An x86 instruction accesses at most $3$
address locations.}. Lets call this as a pigeonhole set. Thus, any subsequent
instructions that access any other page (either code or data) will fall out of
the pigeonhole set resulting in a page fault~\footnote{Note that the process
does not suffer denial of service, only its progress is slowed down due to
excessive page faults.}.  The faulting address of this instruction reveals
what the process is trying to access. In most applications, a large fraction
of memory accesses patterns are defined by the input. To extract the
information about this input,  the OS can pre-empt the process by inducing a
page fault on nearly every instruction.  Our analysis shows that empirically,
every $10$th code / data access crosses page boundaries on an average in
standard Linux  binaries~\footnote{We tested  {\sc CoreUtils} utilities under
random inputs.}. This implies that the OS can single step the \execution at
the granularity of $10$ instructions to make observations about the virtual
address access patterns. Thus, by resorting to this extremity the OS achieves
the maximum leakage possible via the page fault channel.

\begin{figure}
\centering
\epsfig{file=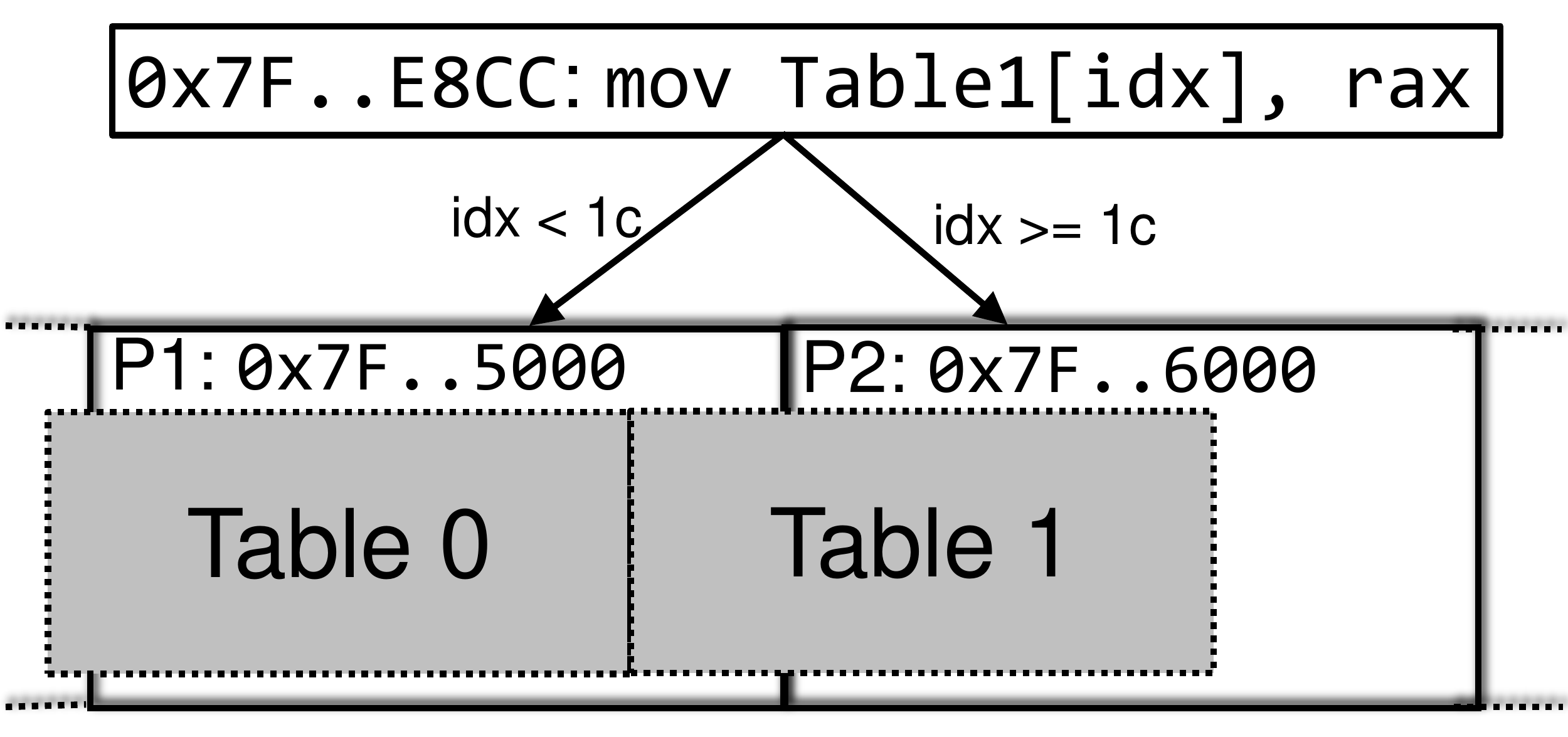, width=8.5cm,height=3.9cm}%scale=0.2}
\vspace{-.5cm}
\caption{Attack via input dependent data page access in AES. 
The data lookup in either in $P_1$ or $P_2$, which is decided by secret byte.}
\label{fig:egdf}
\vspace{-.6cm}
\end{figure}
\subsection{Attack Examples}
\label{sec:example}

A \attack can manifest in any generic application running in an enclaved
environment. In this work, we limit our examples to cryptographic
implementations for two reasons. First, even a minimalistic enclave will at
least execute these routines for network  handshake, session establishment and
so on.  For example, SGX applications such as OTP generators, secure ERM,
secure video conferencing, etc. use an enclave for the TLS connections and
other cryptographic functions on sensitive data~\cite{sgx-apps}.   Second, the
previous work does not study the leakage via page faults in cryptographic
routines since they are assumed to be already hardened against other 
side-channel attacks such as timing and power consumption.  On the contrary, we
show that cache hardening and memory encryption is not enough. 
This is because caches are accessed by lower address bits while
pages are accessed by higher order bits. Only masking lower order bits does not
necessarily mask the page access order. Let us take a look at two
representative examples to demonstrate real \attacks.

% the cache  accesses are fixed based on the program behaviour whereas the page
% tables are under OS's control which gives it more control to influence the
% access patterns. Further, 

% \vspace{-.7cm}
\begin{figure}
\centering
\epsfig{file=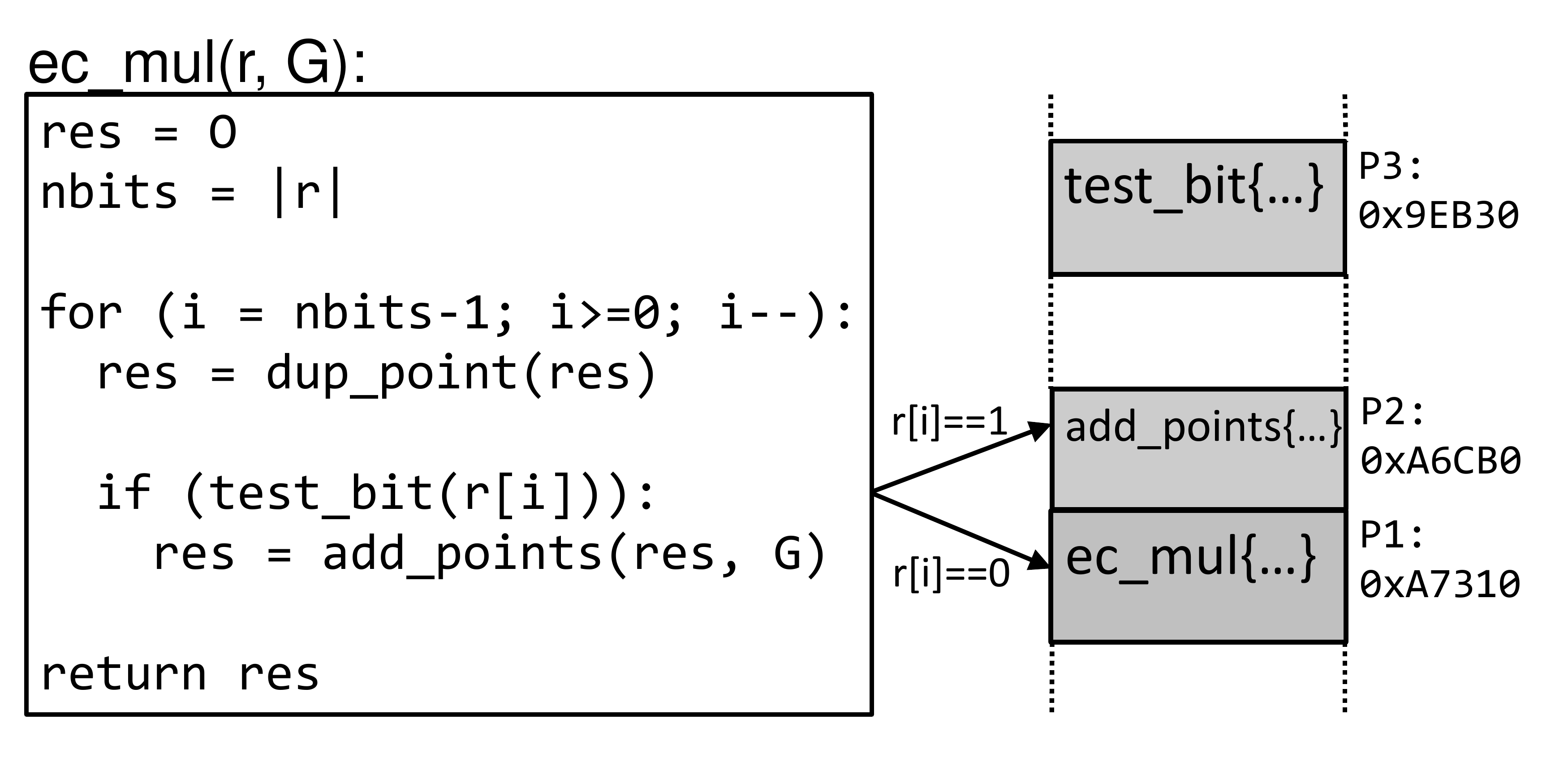, width=8.5cm,height=4cm}%scale=0.2}
\vspace{-0.6cm}
\caption{Attack via input dependent control page access in \eddsa implementation.
The control to either $P_1$ or $P_2$ is dependent on secret bit.}
\label{fig:egcf}
\vspace{-.7cm}
\end{figure}

\paragraph{Input Dependent Data Page Access.} 
We choose a real example of AES from the \libgcrypt v1.6.3 compiled
with \texttt{gcc} v4.8.2 on Linux system. In this example, the adversary can
learn $25$ bits of the input secret key.  Note that the best known purely
cryptanalytic attack for AES leak about 2-3 bits of information about the
key~\cite{aes-biclique}. Any leakage beyond that is a serious amount of
leakage. A typical AES encryption routine involves multiple S-Box lookups.
This step is used to map an input index to a non-linear value, followed by the
MixColumn step~\cite{aes-nist}. In the \libgcrypt implementation
of AES, the lookup tables are designed to contain both S-box values as well as
pre-computed  values for MixColumns transform  for
optimization~\cite{libgcrypt}. There are four such tables ($Table_0$ to
$Table_3$) which are used in table look-ups at various rounds of encryption
process. All the lookup operations in the first round take in a byte of the
secret input key,  XOR it with the plain text (which can be set to 0s) and emit a
corresponding value in the table. Each of these tables comprise of $256$
entries and are statically loaded based on the compiler-generated layout. In
our example,  $Table_1$ and $Table_3$ cross page boundaries. Specifically,
indexes below \texttt{0x1C} are in first page ($P_1$)  while the indexes from
\texttt{0x1C} to \texttt{0xFF} are in second page ($P_2$).
Figure~\ref{fig:egdf} shows the snapshot of the virtual address space of AES,
where $Table_1$ is loaded. During an enclaved execution, the process  will
exhibit \profile depending on the input secret key and the plain text. 
The adversary adaptively selects the plain text and observes the page faults to learn the secret key.
For example,  lets say the key is \texttt{0x1A3E0946} and the adversary choses the plain text to be \texttt{0x00000000}. 
Then the resulting XOR is \texttt{0x1A3E0946}, and the \profile will be  [$P_1 P_2 P_1 P_2$]. An
adversarial OS observing these page faults knows if the enclave is accessing
page $P_1$ or $P_2$. Thus, for each access, this information reduces the OSes
uncertainty from $256$ choices to either $28$ or $228$ choices. In case of
AES, these two portions of the table are accessed $4$ times each in every
round for a 128 / 196 / 256-bit key.  The OS can adaptively execute the
process for different known plain texts and observe the access \profile across multiple runs.  This amounts to
a leakage of $25$ bits in just the first of the total 10 / 12 / 14 rounds of
AES. Thus, $25$ bits is a lower bound. We have experimentally confirmed this leakage (See Appendix~\ref{appx:attacks} for details).

\paragraph{Input Dependent Code Page Access.}
As a second example, consider \eddsa which is an elliptic curve using with
twisted Edward curve and is used in GnuPG and SSL. In \eddsa signing
algorithm~\cite{ed25519},  the main ingredient is a randomly chosen scalar
value $r$ which forms the session key.  The value of $r$ is private and if
leaked it can be used to forge a signature for any arbitrary message.  We show
how the adversary can use \attacks to completely leak  the private value $r$.
Figure~\ref{fig:egcf} shows a code snippet and the page layout for the  scalar
point multiplication routine of \libgcrypt implementation compiled with
\texttt{gcc} v4.8.2.   It takes in an integer scalar ($r$ in this case),  a
point ($G$), and sets the result to the resulting point. The multiplication is
implemented by repeated addition --- for each  bit in the scalar, the routine
checks the value and decides if it needs to perform an addition or not.  The
main routine (\texttt{ec\_mul}), the sub-routines for   duplication
(\texttt{dup\_point}) and testing the bit (\texttt{test\_bit}) are located in
three different pages denoted as $P_1$, $P_2$, $P_3$ respectively.  Interestingly, the
addition sub-routine (\texttt{add\_points}) is located in pages $P_1$ and $P_2$.
A page profile satisfying a regular expression [$P_1$ $P_2$ $P_1$ $P_3$ $P_1$ $(P_1P_2)^{*}$] 
implies a  bit value $1$  and [$P_1$ $P_2$ $P_1$ $P_3$ $P_1$]
implies a $0$ bit value.  Essentially, the OS can learn the exact value of the
random integer scalar $r$ picked by the process. This amounts to a total
leakage of the secret, and in fact enables the OS to forge signatures
on behalf of the enclave.

We demonstrate more attacks on cryptographic implementations of \libgcrypt and
\openssl in Section~\ref{sec:cases}.  These attacks apply to cloud
applications such as multi-tenant web servers and MapReduce platforms~\cite
{paas-attack, vm-cloud, haven, vc3, mr}.

% In this case, the OS can see if there is a transition from the $P_1$
% (\texttt{ec\_mul}) to $P_3$ (\texttt{add\_point}) corresponding to 
% each bit in the scalar.  
%  A transition implies that the bit is $1$ else the
% program will continue the loop and call $P_2$ (\texttt{dup\_points}). 
% Interestingly, the main routine
% (\texttt{\_gcry\_mul\_point}), the sub-routines for testing the bit
% (\texttt{mpi\_test\_bit}) and addition (\texttt{ec\_add\_point}) are located
% in three different pages --- lets refer to them as $P_1$, $P_2$, $P_3$. In this
% case, the OS can see if there is a transition from the $P_1$
% (\texttt{\_gcry\_mpi\_ec\_mul\_point}) to $P_3$ (\texttt{ec\_add\_point}) corresponding to 
% each bit in the scalar.   A transition implies that the bit is $1$ else the
% program will continue the loop and call $P_2$ (\texttt{dup\_points}). This
% results in a regular expression for the \profile where [$P_1$ $P_2$
% $P_1$] implies a $0$ bit value and [$P_1$ $P_2$ $P_1$ $P_3$ $P_1$]  implies a
% $1$ bit value. 
% Note that there is a  one-to-one mapping between the secret
% value and  the \profile. 
% Essentially, the OS can learn the exact
% value of the random integer scalar $r$ picked by the process. This amounts to
% a total leakage of the process secret, and in fact enables the OS to forge
% signatures on the behalf of the enclaves.

% Thus, the OS can observe input-dependent data and control transfers across
% pages, and infer program secrets.  
\section{Overview}
\label{sec:ovw}
The malicious OS can use \attacks to observe the input-dependent memory
accesses and learn the input program secrets. We now discuss our approach to
prevent this leakage.

\begin{figure*}[t]
\centering
	\begin{minipage}[h]{0.23\textwidth}
	\begin{lstlisting}
	foo (int x, int y) 
	{
		z = 2 * y
		if (z != x) 
		{
			if (z < x + 10)
				path_c()
			else
				path_b()
		}
		else 
			path_a()		
	}
	\end{lstlisting}
	\end{minipage}
	\begin{minipage}[h]{0.37\textwidth}
		\epsfig{file=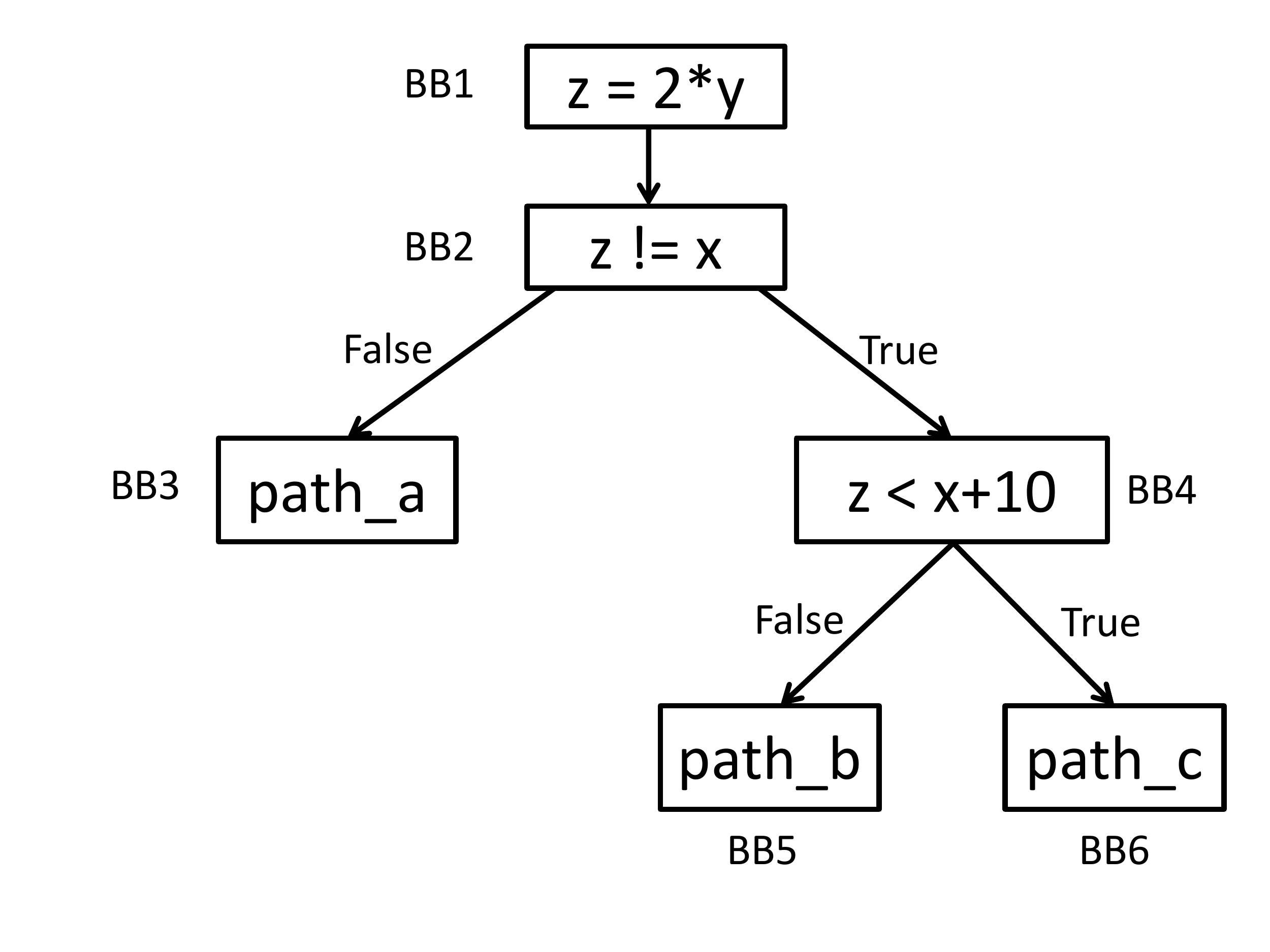, scale=0.25}
	\end{minipage}
	\begin{minipage}[h]{0.34\textwidth}
		\epsfig{file=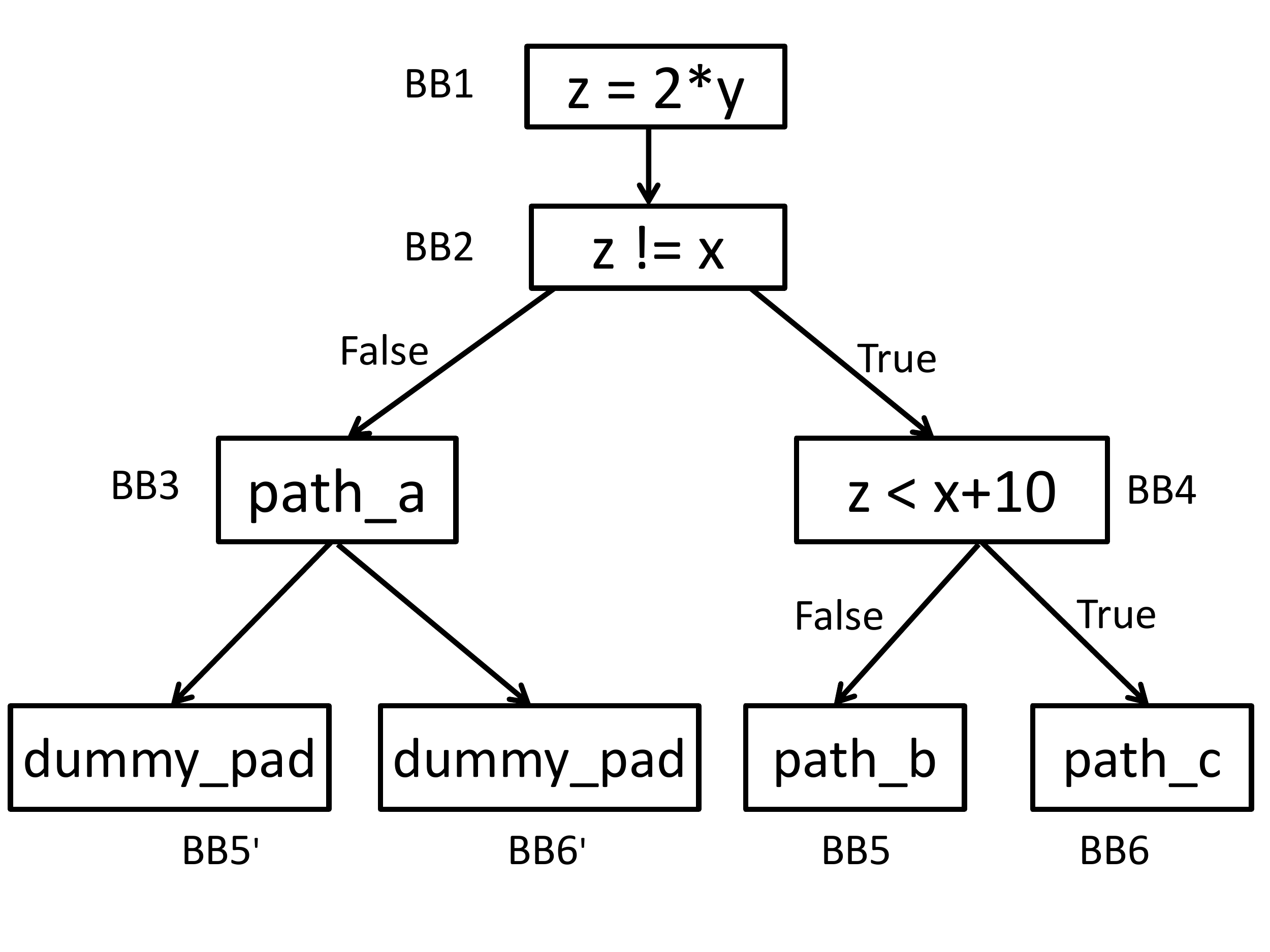, scale=0.25}
	\end{minipage}
\vspace{-.7cm}
\caption{(a) Code snippet for example function \texttt{foo} where \texttt{x} and \texttt{y} are secret. (b) Unbalanced
execution tree. (c) Corresponding balanced execution tree.}
\label{fig:prog-tree}
\vspace{-.6cm}

\end{figure*}

\subsection {Security Definitions \& Assumptions}
\label{sec:def}
Lets represent an enclave program $P$ that computes on inputs $I$ to produce
output $O$ as $(P,I) \mapsto O$, such that both $I$ and $O$ are secret and are
encrypted in RAM.  In case of \execution, the adversary can observe the
sequence of page faults. We term this knowledge of the adversary as the {\em
page access profile}. Note that each observed profile is specific to an input
to the program, and is defined as:

\begin{definition}(Page Access Profile.)
{\em For a given program $P$ and a single input $I$, the page access profile
$\overrightarrow{VP}_{I}$ is a vector of tuples $\langle VP_i \rangle$, where
$VP_i$ is the virtual page number of the  $i^{\text{\tiny th}}$  page fault observed by the OS.} 
\end{definition}

To model the security, we compare the execution of a program on a real
enclaved system with its execution on an ``ideal'' system. The ideal system is
one which has infinite private memory and therefore the program execution
doesn't raise faults. On the other hand, the real system has limited memory and the
enclave will incur page faults during its execution.  Specifically, we define
these two models as follows:

\begin{itemize}
\squish
	\item {$\infty$-memory Enclave Model ($M_{\infty-model}$).}
	The \execution of program on a system with an unbounded physical memory 
	such that the \profile is $\emptyset$.

	\item {Bounded-memory Enclave Model ($M_{B-model}$).}
	\Execution of program such that for any instruction in
	the program, the enclave has the least number of pages required for
	executing that instructions~\footnote{In our case it is at most three
	pages, which is the maximum number of pages required to execute any Intel
	x86 instruction.}. 
	% We assume that the OS can take away the enclave pages only
	% when handling a page fault.
	
\end{itemize}

\begin{definition}(Page Access Profile Distinguishability)
{\em Given a program $(P, {I}) \rightarrow {O}$, we say $P$ exhibits
\profile distinguishability if there exists an efficient adversary $\cal{A}$ such that 
$\exists~I_0, I_1 \in {I}$ and $b \in \{0, 1\}$, for which the advantage:% defined as 
\\
Adv$(\cal{A}) = |$Pr$[$Exp$(\overrightarrow{VP}_{I_{b = 0}}) = 1]$ $-$ Pr$[$Exp$(\overrightarrow{VP}_{I_{b = 1}}) = 1]|$
\\ is non-negligible.} 
\end{definition}

If a probabilistic polynomial time-bounded adversary can distinguish the
execution of the program for two different inputs by purely observing the
\profile, then the program exhibits \profile distinguishability. A safe
program exhibits no leakage via the page fault channels; we define page-fault
obliviousness as a security property of a program as follows:

\begin{definition}(PF-obliviousness)
{\em Given a program $P$ w.r.t. inputs $I$, the PF-obliviousness states that
if there exists an efficient adversary $\cal{A}$ which  can distinguish
$(\overrightarrow{VP}_{I_0}, \overrightarrow{VP}_{I_1})$ for $\exists~I_0, I_1
\in \vec{I}$ in the $M_{B-model}$, then there exists an adversary $\cal{A^\prime}$
which can distinguish ${I_0}$, ${I_1}$ in the $M_{\infty-model}$.} 
\end{definition}

Our definition is a relative guarantee --- it states that any information that
the adversary learns by observing the execution of program on a bounded
private  memory, can always be learned by observing the execution even on an
unbounded memory (for e.g., the total runtime of the program). Such
information leaked can be gleaned even without the page fault channel. Our
defense does not provide any absolute guarantees against all possible side-
channels. If there are additional side channels in a  PF-oblivious program,
they can be eliminated with orthogonal defenses.

\paragraph{Scope and Assumptions.}
Our work considers a software-based adversary running at ring-0; all 
hardware is assumed to be trusted. Further, the following challenges are 
beyond the goals of this work: 
\begin{itemize}
\squish

	\item {\bf A1.} Our attacks and defenses are independent of other side-channels such as
	time, power consumption, cache latencies, and minor execution time differences
	between two different memory access instructions that raise no faults.   If
	such a difference is discernible, then we can show that they provides a
	source of advantage even in an execution with no page faults ($\infty$-model).
	Application developers can deploy orthogonal defenses to prevent against these
	side-channels~\cite{dangfeng-lang}. Our defenses do not prevent information
	leakage via untrusted I/O, system-call, and filesystem channels~\cite{iago}.

	\item {\bf A2.} Once a page has been allocated to the enclave, the OS can take it away	
	only on a memory fault. We do not consider the case where
	the OS removes enclave pages via a timer-based pre-emption, since the
	adversary's  clock granularity is much coarser in this case and likely yields
	a negligible advantage.

\end{itemize}

\subsection{Problem \& Approach Overview} 
\label{sec:soln}

\paragraph {Problem Statement.} Given a program $P$ and  set of secret inputs $I$,
we seek a program transformation $T$: $P \mapsto P^\prime$ such that the
transformed program $P^\prime$ satisfies \property with respect to all
possible values of $I$.

% If we can transform a program so as to guarantees oblivious execution, then
% our transformation function is safe with respect to the page fault side
% channel. Thus, if the \profile of the \execution is same irrespective of the
% input, the observer will not learn anything about the sensitive input.

Consider a program executing on sensitive input. The execution path of such a
program  can be defined by the sequence of true and false branches taken at
the conditional statements encountered during the execution. Each set of
straight-line instructions executed and corresponding data accessed between
the branching condition statements can be viewed as an {\em execution block}.
Let us assume that each execution block has the same number of memory accesses
and by assumption A1 each  memory access takes approximately same amount of
time.  Then, all such paths of a program can be represented using a tree, say
the {\em execution tree} such that each node in the tree is an execution block
connected by branch edges.  For example, the function \texttt{foo()} in
Figure~\ref{fig:prog-tree}~(a) has $3$ execution paths in the execution tree
shown in Figure~\ref{fig:prog-tree}~(b). Each of the paths   a, b, c can be
executed by running the program on the inputs   (x = 4, y = 2), (x = 8, y = 9)
and (x = 6, y = 5) respectively.

\Profile is inherently input dependent, so anyone who observes the \profile
can extract bits of information about the input. However, if the \profile
remains the same irrespective of the input, then the leakage via page fault
channel will drop to zero~\cite{oblivm, csf13}.  We call this transformation
strategy as determinising the \profile. We adopt this strategy and enforce a
deterministic \profile for possible paths in the program execution. The
\execution always sequentially accesses all the code and data pages that can
be used at a particular memory-bound instruction for each execution.  In our
example, Figure~\ref{fig:prog-tree}, we will  access both BB3 as well as BB4
irrespective of the branching condition.  Similarly, we also apply it at level 4,
so that the complete program path is {BB1, BB2, BB3, BB4, BB5${^\prime}$, BB6${^\prime}$, BB5, BB6} for all
inputs.  Thus, deterministic execution makes one real access and several fake
accesses to determinise the \profile. It is easy to see that under any input 
the execution exhibits the same \profile.

The challenge that remains is: how to execute such fake accesses while still
doing the actual intended computations. We present a simple mechanism to
achieve this. First we use the program's execution tree to identify what are
all the code and data pages that are used at each level of the tree for all
possible inputs (BB3, BB4 at level 3 in our example).  This gives us the set of
pages for replicated-access.  Next, we use a multiplexing mechanism to 
load-and-execute the correct execution block. To achieve this, we break each code
block execution into a fetch step and an execute step. In the fetch step, all the
execution blocks at the same level in the execution tree are fetched from memory 
sequentially. In the execute step the multiplexer will select the real block
and execute it as-is.  In our example, for (x = 4, y = 2), the
multiplexer will fetch all blocks but execute only BB3 at level 3,  and for (x = 8, y = 9)
or (x = 6, y = 5), the multiplexer will execute BB4.

\section{Design}
\label{sec:design}
% There are multiple transformations which can be used to achieve \property for a given program. 

There can be several ways for determinising the \profile; selecting the best
transformation is an optimization problem. We discuss one such transformation
which can be applied generically and then present the program-specific
transformations which incur lower costs (Section~\ref{sec:optimization}).

\subsection{Setup}

It is simple to adapt the standard notion of basic blocks to our notion of
execution blocks. In our example code snippet in Figure~\ref{fig:prog-tree}~(a), 
 we have $6$ such execution blocks BB1 to BB6. In case of BB1, the
code page $\mathbb{C}$ will comprise of virtual page address of the statement
\texttt{z = 2 * y}, and data pages $\mathbb{D}$ will have virtual page address of
variables \texttt{z} and \texttt{y}.

Note that the execution tree in Figure~\ref{fig:prog-tree}~(b) is unbalanced,
i.e., the depth of the tree is not constant for all possible paths in the
program. This imbalance in itself leaks information about the input to an
adversary even without \attacks simply by observing the function start-to-end
time.  For example, the first path (\texttt{path_a}) in Figure~\ref{fig:prog-tree}~(b) 
  is of depth $2$ and is only taken when value of z equals value of
x. If the adversary can try all possible values of secret, then the tree depth
becomes an oracle to check if the guess is correct. To capture the information
leaked strictly via the page fault channel, we limit our scope to balanced
execution tree. If the tree is unbalanced, then the input space is partitioned
into sets which are distinguishable in the original program in the
$\infty$-model. Since we limit our scope to achieving indistinguishability
relative to  $\infty$-model, we safely assume a balanced execution tree as
shown in Figure~\ref{fig:prog-tree}~(c)~\cite{pc-security}. Techniques such as
loop unrolling, block size balancing with memory access and \texttt{NOP}
padding can be used to balance the tree depth and block sizes~\cite{coppen}.
In our experience, cryptographic routines which are hardened against timing
and cache side-channels generally exhibit balanced execution trees. For the
set of programs in our study, if necessary, we perform a 
pre-preparation step manually to balance the execution tree explicitly.

% \vspace{-.1cm}
% \begin{definition}(Execution Block.)
% {\em A execution block B represented as ($\mathbb{C}$, $\mathbb{D}$), where
% $\mathbb{C}$ is an ordered set of all the code pages and $\mathbb{D}$ is the
% corresponding data pages accessed by each instruction in $\mathbb{C}$.}
% \end{definition}
% % \vspace{-.2cm}

% % \vspace{-.1cm}
% \begin{definition}(Execution Tree.)
% {\em 
% An execution tree $T = \langle V, E \rangle$ where each vertex $V$ is an execution block
% connected by conditional edge $E$ to form an execution tree.}
% \end{definition}
% % \vspace{-.2cm}
% % \vspace{-.1cm}
% \begin{definition}(Balanced Execution Tree.)
% {\em Each program path has equal depth of execution blocks, and each execution block
% takes same amount of time to execute.}
% \end{definition}
% % \vspace{-.2cm}
% A program fragment with balanced execution tree depth of $n$, is such that
% each path in the program has $n$ execution blocks.  

Even after the execution tree is balanced, the pigeonholing adversary knows
the sequence of the execution blocks that were executed for a given input only by
observing page faults. For example, lets assume that the execution blocks BB5 and
BB6 are in two different pages $P_1$ and $P_2$ respectively. Then the result
of the branching condition $z < x + 10$ will either cause a page fault for
$P_1$ or $P_2$, revealing  bit of information about the sensitive input
\texttt{x} and \texttt{y}. Given a balanced execution tree, we design 
a transformation function to make the \profile independent of the input~\cite{csf13}. 

\subsection{\Interpreting}
\label{sec:interpreting}
We now discuss a concrete design of our transformation namely \interpreting and 
demonstrate how it can be supported to transform legacy C / C++ applications in the 
current compiler infrastructure.

\paragraph{Basic Multiplexing.}
In the fetch phase, we copy the code blocks at the same level of the execution
tree  to a temporary page --- the code staging area ($SA_{code}$). All data
that may be used by each of these sensitive code blocks is copied to a
separate temporary page --- the data staging area ($SA_{data}$). Then in the
execution phase, we use an access multiplexer which selects the correct code
and data blocks and executes it (by jumping to it).  At the end of the
sensitive execution, the content from data staging area is then pushed  back
to the actual addresses. If the execution changes any data in the staging area, the new values are
updated. The rest of the values are just copied back unchanged. Note
that all these operations are done in a sequence in the staging area (one code page). 
Thus this execution is atomic --- no page faults can occur between them. 
From an adversarial viewpoint, the execution is performed
within the boundary of single code and single data page. 
So all that the adversary can see is the same sequence of page faults for any input.
Thus our multiplexed fetch and execute mechanism ensures that the OS cannot determine which code and data block was actually used within the staging area.

\begin{figure}
\centering
\epsfig{file=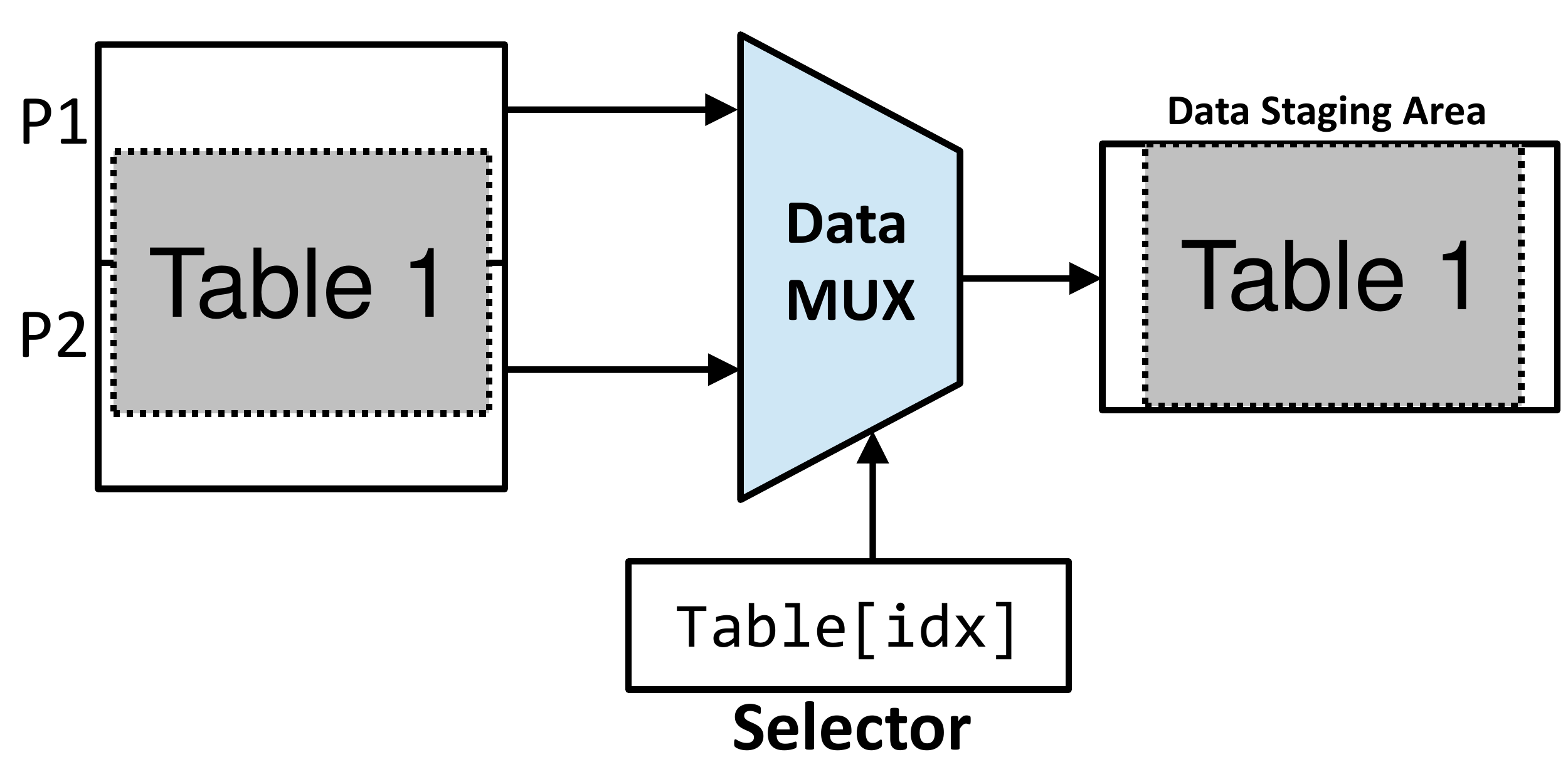, width=7cm, height=3cm} %scale=0.2]
\vspace{-.35cm}
\caption{\Interpreting to prevent leakage via data access. 
The multiplexer accesses the correct offset in the staging area.}
\vspace{-.4cm}
\label{fig:data-example}
\end{figure} 

\begin{figure}
\centering
\epsfig{file=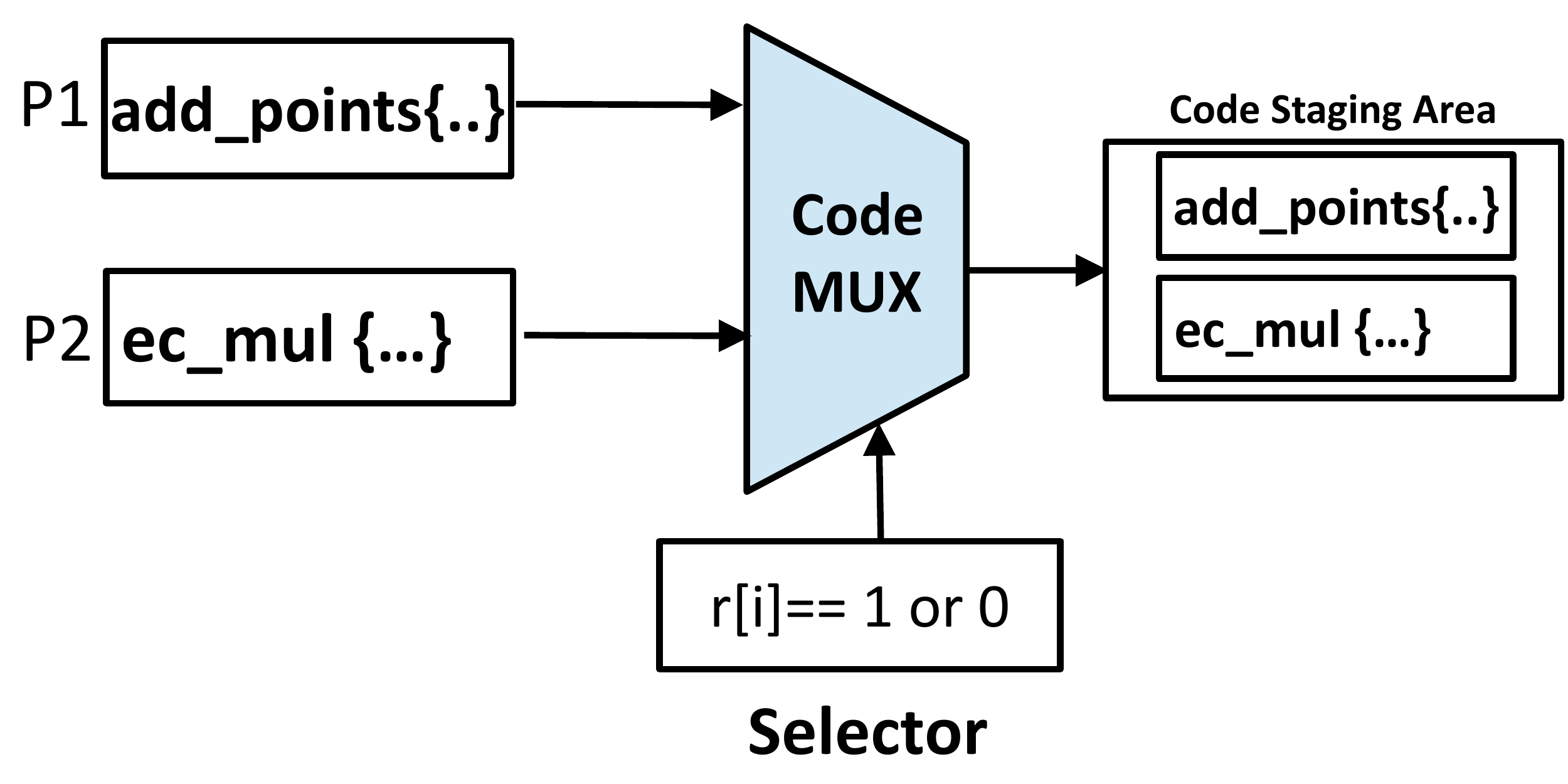, width=7.5cm, height=3cm}
\vspace{-.35cm}
\caption{\Interpreting to prevent leakage via code page access. 
The multiplexer executes the correct function in the staging area.}
\vspace{-.6cm}
\label{fig:code-example}
\end{figure}

\paragraph{Example.} 
For our AES case, we apply \interpreting and copy the  data
table $T_3$ to staging area (See  Figure~\ref{fig:data-example}). Each data
access now incurs $2$ data page copies and a code page copy followed by
multiplexed accesses.  Similarly for \eddsa, we can multiplex 
the called functions into $SA_{code}$ (See  Figure~\ref{fig:code-example}). 
This asserts that the OS cannot differentiate  whether the true or
the false branch was executed, by looking at the \profile. Thus, in both the
cases the OS can observe the fetch and execute operations only at the page
granularity. It cannot determine which of the fetch or execution operations is
real and which is replicated.

\paragraph{Compacted Multiplexing.}
In the multiplexing mechanism, it is important that both $SA_{code}$  and
$SA_{data}$ must fit in a single page each to prevent information leakage. For
ensuring this, we specifically pick a block size such that at any given level
in the execution tree, all the blocks and the corresponding data always fit in
a single page. However, there are cases where the execution tree is deep and
has large number of blocks (total size of more than $4096$ bytes) at a certain
level. This results in a multi-page staging area. To address this, we  use a
compaction scheme to fit the staging area in a single page.  Specifically, in
the fetch phase we create a dummy (not real) block address  in the staging
area. The blocks which are not going to be executed are  saved at this dummy
location during the fetch step. Each new block from the execution tree
overwrites (overlap) the same location. Only the real block (which will be
executed) is copied in a non-overlapping address in the page. We term this as
a {\em smart copy} because each copy operation writes to either dummy or real
page-offset in the staging area. The adversary OS does not see the offset of
the faulting address,  and hence cannot distinguish a dummy vs. a real copy.
Thus the staging area always fits in a single page. The semantics of the execute phase are unchanged.

% copy to the same offset vs. a copy to a different offset within the same page. 
% In this way, we compact and fit all the execution blocks into a 
% single block and the real block is copied to a separate location within the staging area.

% The OS can
% only observe at the granularity of a page (and not the offsets), hence this is
% safe. 

\subsection{Compiler-enforced Transformations}
\label{sec:automation}

\begin{figure}
% \begin{center}
\scriptsize
\begin{tabular}{llll}
                
Labels      & $\mathcal{L}$ & ::= & \texttt{$high$ | $low$} \\\\

Expressions & e    & ::= & e$_{1}$ $\oslash$ e$_{2}$ $|$ e$_{1}$ ? e$_{2}$ : e$_{3}$ $|$ $\circledcirc$ e\\
            &      & ::= & $|$ $\odot$\texttt{lval} $|$ \texttt{lval} $|$ c $|$ \\
            &      & ::= & $|$ \texttt{foo} (e$_{1}$, e$_{2}$, $\mathellipsis$, e$_{n}$) \\
            & c    & ::= & \texttt{const} \\
            & lval & ::= & \texttt{var} $|$ \texttt{var} [e] \\\\
           
Unary & $\odot$         & ::=   & \& $|$ * $|$ $++$ $|$ $--$ \\
      & $\circledcirc$  & ::=   & $\mathtt{\sim}$ $|$ ! $|$ $+$ $|$ $-$ $|$ \texttt{sizeof} \\\\
                 
Binary & $\oslash$ & ::= & $+$ $|$ $-$ $|$ * $|$ / $|$ \%  \\
       &                &     & $|$ \&\& $|$ $>>$ $|$ $<<$ \\
       &                &     & $|$ $|$ $|$ \& $|$ \char`\^{ }  $|$ != $|$ == \\
       &                &     & $|$ $>$ $|$ $<$ $|$ $>$= $|$ $<$= \\\\

Commands & P & ::= & lval := e \\
        &   &     & $|$ \texttt{if} (e) \texttt{then} P \texttt{else} P'\\
        &   &     & $|$ \texttt{do} \{P\} \texttt{while} ($\lceil$e$\rceil^{c}$) \\
        &   &     & $|$ \texttt{while} ($\lceil$e$\rceil^{c}$) \texttt{do} \{P\} \\
        &   &     & $|$ \texttt{for} (e$_{1}$ ; $\lceil$e$_{2}$$\rceil^{c}$ ; e$_{3}$) \{P\} \\
        &   &     & $|$ \texttt{foo} (e$_{1}$, e$_{2}$, $\mathellipsis$, e$_{n}$) \{P\} \\
        &   &     & $|$ \texttt{return} e \\\\

Program & S & ::= & \texttt{begin\_pf\_sensitive} P \\
        &   &     & \texttt{end\_pf\_sensitive} \\\\

\end{tabular}
\vspace{-.6cm}
\caption{The grammar of the language supported by our compiler. $\lceil$e$\rceil^{c}$ denotes that the loop is bounded by a constant $c$.}
\label{fig:grammar}
% \end{scriptsize}
\vspace{-.6cm}
% \end{center}
\end{figure}
% \squish

We build our design into the compiler tool chain which works on a subset  of C
/ C++  programs. Figure~\ref{fig:grammar} describes the mini-language
supported by our compiler which can transform existing applications. Given a
program, the programmer manually annotates the source code to demarcate the
secret input to the program and specifies the size of input with respect to
which the transformation should guarantee \property. Specifically, he manually
adds compiler directive \texttt{begin\_pf\_sensitive} and
\texttt{end\_pf\_sensitive} to mark the start and end of sensitive code and
data.  For example, the developer can mark the encryption routine, decryption
routine, key derivation, key, nounce, and so on as secret.  Our tool comprises
of analysis and transformation steps to enforce \interpreting which are
discussed next.

\paragraph{Identifying Sensitive Code and Data.}
In the first step, our compiler front-end parses the source code and
identifies the programmer added directives. It then performs a static analysis
which transitively marks all the instructions and variables within the
lexical scope of programmer-marked sensitive code as \texttt{high}. 
Non-sensitive instructions and variables are marked as
\texttt{low}. At the end of the phase, each instruction and variable in the
code has a sensitivity tag (\texttt{high} or \texttt{low}).

% Table~\ref{tbl:transforms} summaries the static label propagation rules used to propagate the
% \texttt{high} and \texttt{low} sensitivity tags for various types of
% operations in the program. 

\paragraph{Determinising the Page-layout.}
Next, our tool performs an analysis to decide the new virtual address layout
for the sensitive data and code (marked as \texttt{high})  for placing them in
the staging area. The initial step is to identify  the existing execution tree
of the sensitive code. To achieve this,   we create a super-CFG wherein each
function call is substituted with the body of the function and all the bounded
loops are unrolled. This creates an execution tree such that all the sensitive
execution blocks are identified. We seek a mapping $\Gamma:
B \mapsto \mathcal{L}$ such that  all the
execution blocks at the same level in the execution tree are relocated to the
same virtual page address. There are multiple possible $\Gamma$ mappings which
yield acceptable layouts,  but our goal is to select the one where the code
and data staging areas always fit in a single page. We first try to use the
basic multiplexing for arranging the blocks if the total size of all the
blocks at a level is less than $4096$ bytes.  If the size of the required
staging area exceeds one page, then we resort to compacted multiplexing (See
Section~\ref{sec:interpreting}).

% where $B$ is the set of execution blocks and $\mathcal{L}$ is an ordered-list of 
% corresponding re-located virtual page addresses. The execution blocks are relocated such that $\forall b_{i,j} \in B_j$, 
% where $B_j$ is set of all execution blocks at level $j$ in the execution tree, $\Gamma[b_{i,j}] = L_j$. 

\paragraph{Instruction Rewriting.} 
The last step of transformation comprises of: (a) Adding logic for
multiplexing (b) Adding prologue-epilogue before and after the multiplexing to
move the code / data to and from staging area. Next, we rewrite the
instructions to introduce replicated accesses to data pages, and instrument
each execution block with a call to the code multiplexing logic as described
in Section~\ref{sec:interpreting}. Finally, we add prologue and epilogue
before and after each execution block at each CFG level.

% \paragraph{Preprocessing.}
% Our LLVM tool-chain comprises of alias analysis, loop analysis and CFG
% analysis over the intermediate code to assist our transformation. We build on
% top of \texttt{basicaa}, \texttt{da, memdep}, \texttt{loops}, \texttt{basiccg,
% dot-cfg}, and \texttt{loop-unroll} analyses from LLVM. The loop bounds of our
% input program are determined  statically. Our analysis first tries to  infer
% the correct loop bounds statically, and if is unsuccessful reports an error to
% the programmer. To resolve this, the programmer can choose to annotate the
% input program and specify the loop bounds explicitly. Once resolved, we unroll
% all the loops in the sensitive code. In several case studies, we also require
% to resolve pointers and aliases. We use the flow-sensitive points-to analysis
% algorithm proposed by Hardekopf \etal to resolve necessary pointer addresses
% conservatively~\cite{ben-hardekopf}.  Developers can also refine the imprecise
% results of the points-to inference by providing the points-to information
% explicitly.

\paragraph{Example.} 
In case of \eddsa, we manually add compiler pragmas to mark the user key
variable  and the signing routine as sensitive.  Our analysis phase identifies
$31$ functions, $701$ execution blocks, $178$ variables as
sensitive. It also collects information about the call graph, function CFG and  
access type (read or write) of the variables. After the
analysis, our tool calculates (a) the staging area to be created in first
function \texttt{ec_mul} just before the first access to the key (b) layout of
the data staging area such that all the variables fit in one page (c) the
alignment of the execution block in the staging area, (d) the new addresses of the
sensitive variables used in these execution block, and (e) instructions which are
to be updated for accessing the staging area. Finally, we add code for preparing the staging area and
instrument the code  instructions to use the data staging area values.

\paragraph{Security Invariant.}
The above compiler transformation ensures that for the output program, 
all the execution blocks at the same level in the execution tree are mapped 
to same ordered list of virtual address locations. Thus for all the inputs,
the program exhibits the same \profile hence satisfying our \property property.

\section{Developer-Assisted Optimizations}
\label{sec:optimization}

Apart from the automated-transformation, there can be other 
strategies which have been manually confirmed to make programs PF-oblivious. 
We discuss such performance optimizations which allow 
developer assistance. In the future, our compiler can be extended
to search and apply these optimization strategies automatically.
\begin{figure*}[t]
\centering
	\begin{minipage}[h]{0.3\textwidth}
		\epsfig{file=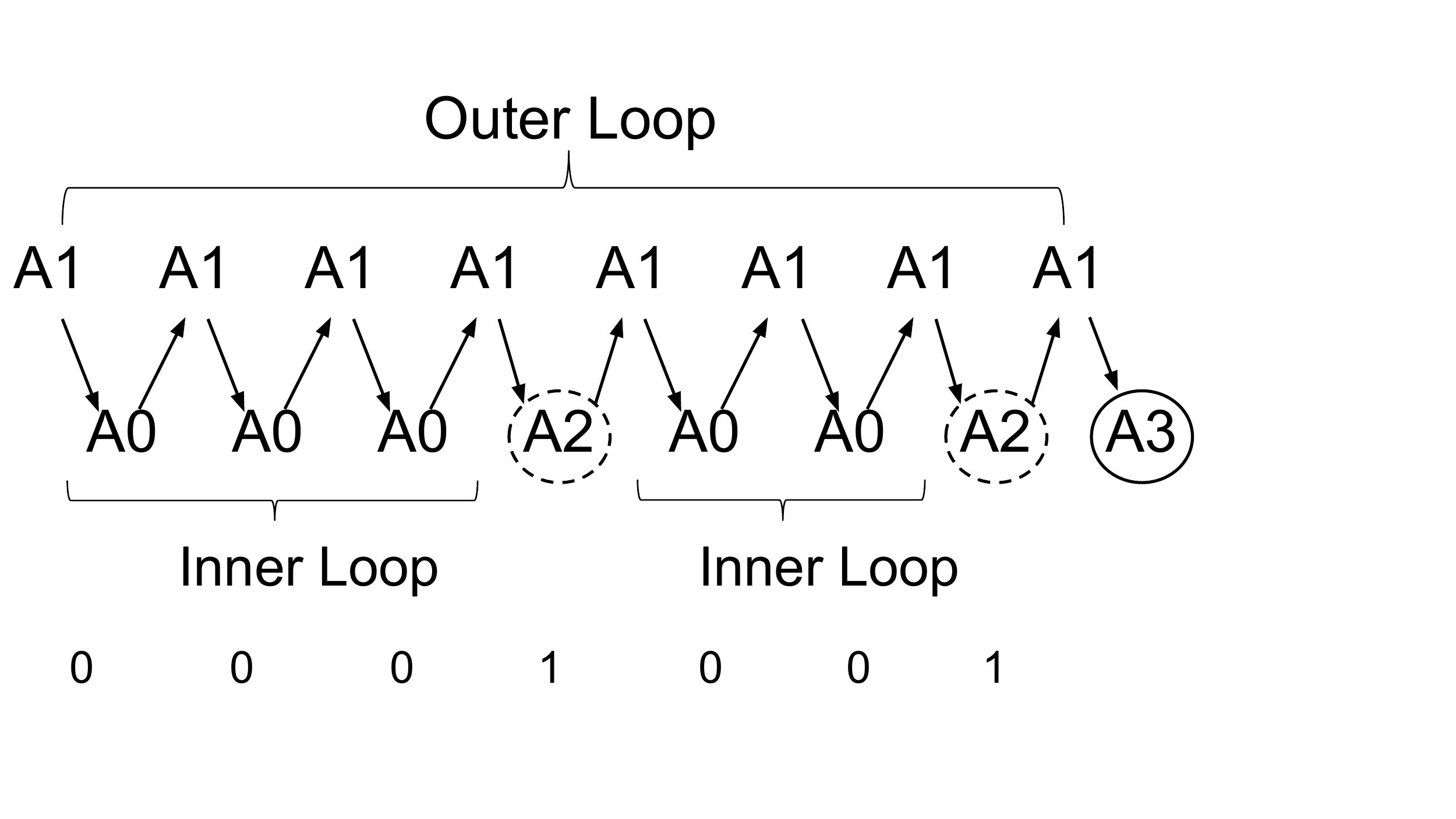, scale=0.26}
	\end{minipage}
	\begin{minipage}[h]{0.34\textwidth}
		\epsfig{file=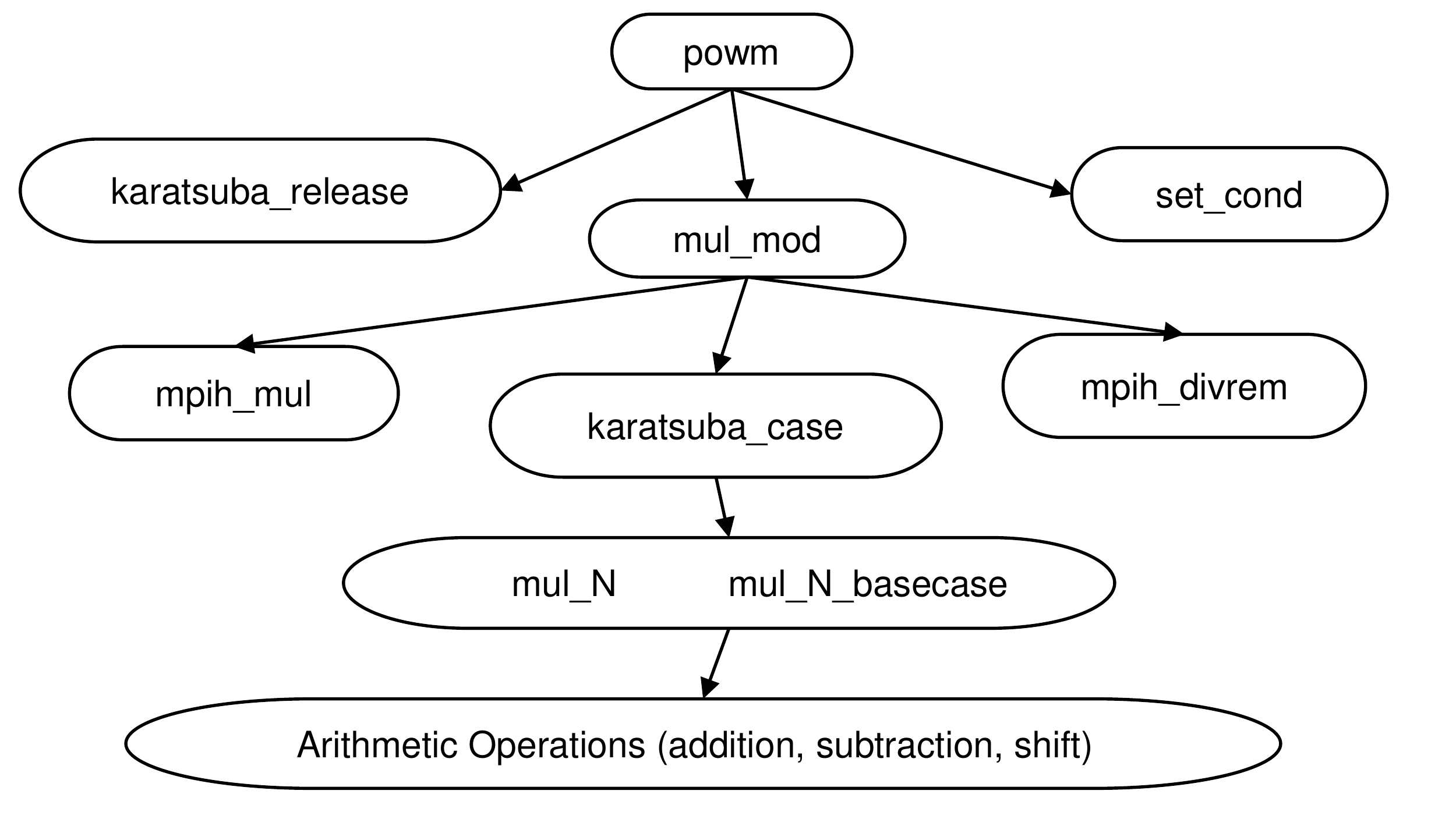, scale=0.25}
	\end{minipage}
	\begin{minipage}[h]{0.32\textwidth}
		\epsfig{file=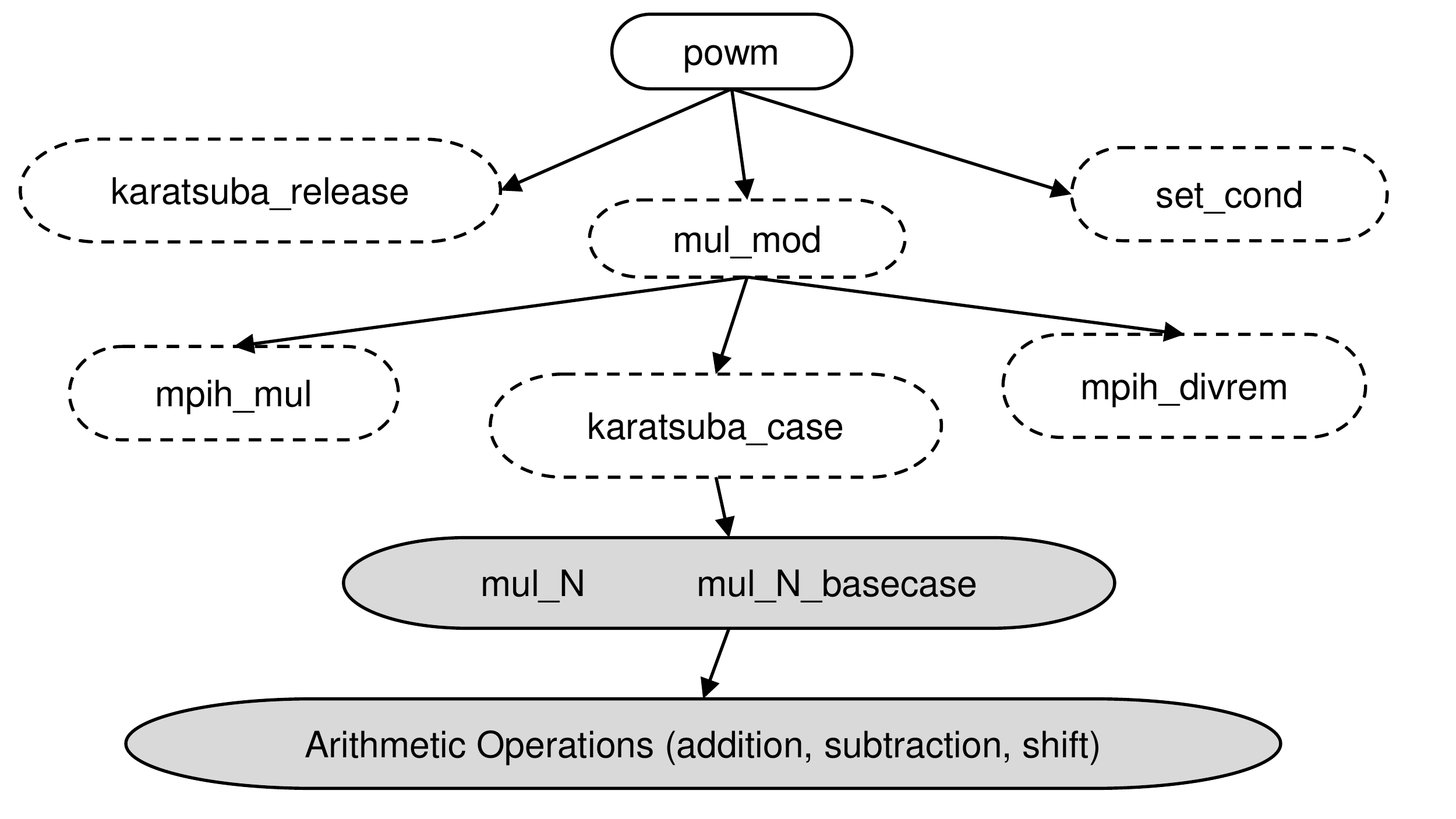, scale=0.25}
	\end{minipage}
\vspace{-.5cm}
\caption{(a) Simplified page access profile for \powm (Window size = 1) where A0, A1, A2, A3 sub-patterns denote transitions between 
functions mul\_mod(),  powm(), set\_cond() and karatsuba\_release() respectively. (b) Call graph before enforcing \interpreting. (c) Alignment after enforcing \interpreting with optimization (O4), dotted and shaded functions are moved to separate code staging pages.}
\label{fig:powm-example}
\vspace{-.5cm}
\end{figure*}

\subsection{Exploiting Data Locality}
The main reason that input-dependent data accesses leak information in
\attacks is that the data being accessed is split across multiple pages.  In
all such cases,  the \interpreting repetitively copies data to and fro between
the staging area and the actual data locations. There are two key observations
specific to these cases. 

\paragraph {O1: Eliminating copy operations for read-only data.} 
We observe that most of the table lookup operations are on pre-computed data
and the code does not modify the table entries during the entire execution.
Since these sensitive data blocks are used only in read operations,  we can
fetch them into $SA_{data}$ and discard them after the code block executes.
This saves a copy-back operation per code block. Moreover, if the next code
block in the execution tree uses the same data blocks which already exist in
$SA_{data}$, then we need not copy them to $SA_{data}$.  This save all the
copy operations after the data is fetched into the $SA_{data}$ for the first
time.  In case of AES, we require only two operation to copy $Table_{1}$ from
$P_1$ and $P_2$ to $SA_{data}$. We can apply the same strategy to $Table_{3}$,
so that the entire execution needs only four copy operations.

\paragraph {O2: Page Realignment.} 
All the data blocks which are spread across page boundaries (specifically,
S-Boxes) can be grouped together and realigned at the start of the page.  This
ensures that the set of sensitive data pages is minimum for the entire
execution.  In the context of AES example, both $Table_{1}$ and $Table_{3}$
cross the page boundary and use $3$ pages.  They can be aligned to page
boundary and fit in $2$ pages. Thus for \interpreting, the patch will incur
only two copy operations in total.

% Note that both of them  do not affect the data or
% code access pattern or the page fault sequences enforced by our compiler  and
% hence they still preserve the security invariant in the transformed
% code.
Note that the above strategies are safe and respect the security invariant (Section~\ref{sec:automation}) because all the
eliminations are  independent of the input and thus the reduction in the copy
operations affects all the inputs uniformly. 
% Specifically,  if $\forall I_0,
% I_1 \in {I}$,  the profile for $\Gamma: B \mapsto L$ is $L_0, \ldots ,L_k$,
% then after the transformation $\forall I_0, I_1 in {I}$,  the profile for
% $\Gamma^\prime: B \mapsto L$ is $L_0, \ldots ,L_k^\prime$, where $k < k^\prime$.

\subsection{Exploiting Code Locality}
In case of input-dependent control transfers, automatically determinising the
control flow results in a high number of multiplexing operations.  To address
this short-coming we propose a set of strategies specific to the type of
pigeonhole attacks, which reduces the overheads to an acceptable range. We
take the example of \powm and demonstrate our  strategies.

\setlength{\textfloatsep}{6pt}
\begin{algorithm}[t]
\scriptsize
\label{sliding_window}
\caption{\libgcrypt modular exponentiation (\powm).}
{INPUT:} {Three integers $g$, $d$ and $p$ where $d_{1}...d_n$ is the binary representation of $d$.} 
{OUTPUT:} {$a \equiv \ {g^d}~(mod~p)$.} 
\begin{algorithmic}
  \Procedure{\powm}{$g, d, p$} \Comment {P1}
    \State $w \gets$ GET_WINDOW_SIZE($d$), $g_0 \gets 1, g_1 \gets g, g_2 \gets g^2$
    \For {$i \gets 1$ to $2^{w-1} - 1$} \Comment {Precomputation}
        \State $g_{2i+1} \gets g_{2i-1}  \cdot g_2$ \texttt{mul_mod} $p$ 
    \EndFor
    \State $a \gets 1$, $j \gets 0$
    \While{$d \neq 0$} \Comment{Outer loop}
        \State $j \gets j +$ COUNT_LEADING_ZEROS($d$)
        \State $d \gets$ SHIFT_LEFT($d, j$)
        \For {$i \gets 1$ to $j + w$} \Comment{Inner Loop}
            \State $a \gets a \cdot a$ \texttt{mul_mod} $p$ \Comment{P2}
        \EndFor
        \State $t \gets d_{1}...d_w$; 
        \State $j \gets$ COUNT_TRAILING_ZEROS($t$)
        \State $u \gets$ SHIFT_RIGHT($t, j$)
        \State $g_u \gets$ FETCH_POWER(\texttt{set_cond}($u$)) \Comment{P3}
        \State $a \gets a \cdot g_u$ \texttt{mul_mod} $p$ \Comment{P2}
        \State $d \gets$ SHIFT_LEFT($d, w$)
    \EndWhile
    % \For{$i \gets 1$ to $j$}
    %     \State $a \gets a \cdot a$ \texttt{mul_mod} $p$
    % \EndFor
    % \State CLEANUP(), return $a$ \Comment {P4};
\EndProcedure
\end{algorithmic}
\label{alg:powm}
\end{algorithm}

Algorithm~\ref{alg:powm} shows the code structure and data access pattern for
the \powm example. In the \libgcrypt implementation, the actual function body
(\powm), the multiplication function (\texttt{mul\_mod}) and the table lookup
function (\texttt{set\_cond}) are located in three separate pages say $P_1$,
$P_2$, $P_3$ respectively. Hence, the leakage from \powm is due to the
different fault patterns generated from calls to \texttt{mul\_mod} and
\texttt{set\_cond} functions.  Figure~\ref{fig:powm-example} (a) shows the
page fault pattern for \powm with respect to these functions and Figure~\ref{fig:powm-example} (b) shows the function arrangement for \powm. Let us consider
the  implementations of \interpreting in Section~\ref{sec:automation} that make calls to both these
functions indistinguishable. For this, we generate the call graphs of both
functions which identifies the set of sensitive functions are to be masked.
For each call to any of these sensitive function, we perform a multiplexing
operation. It iterates over the set of these sensitive functions in a
deterministic manner and copies all the blocks to $SA_{code}$. The multiplexer
then selects the correct block and executes it. In case of \powm, we move
\powm, \texttt{mul\_mod} and \texttt{set\_cond} to the staging area.  This
implementation of Section~\ref{sec:automation} incurs an overhead of $4000\times$, which is
prohibitive. We discuss our strategies in the context of this example to
describe the reasoning for the optimization. %Next, we explain our optimizations.

\paragraph{O3A: Level Merging.} 
The dominating factor in the \interpreting is the number of copy and
multiplexing operations at each level in the execution tree. We observe that
by the virtue of code locality, code blocks across multiple levels can be
merged together in a single level. Specifically, we place the code blocks such
that the caller and callee function are contained within a page.  
% This reduces the number of copy operations since both the blocks are already in the staging area. 
For example, consider 3 code blocks \texttt{a}, \texttt{b}, \texttt{c} located in three separate
pages.  The call graph is such that \texttt{c} is called by both \texttt{a} and \texttt{b}.  If total
size of \texttt{a}, \texttt{b}, \texttt{c} put together is less than a page ($4096$ bytes), then we can
re-arrange the code such that all three of them fit in a single page. In terms
of the execution tree, it means that we fold the sub-tree to a single code
block. 

\paragraph{O3B: Level Merging via Cloning.} 
The above strategy will not work in cases where the code blocks in a sub-tree
cannot fit in a single page.  To address this, we use code replication i.e.,
we make copies of shared code block in multiple pages.  In our example, if 
blocks \texttt{a}, \texttt{b}, \texttt{c} cannot fit into a single page, we rearrange and replicate the block \texttt{c} in
both P2 and P3. After replication, a control-flow to \texttt{c} from neither \texttt{a} nor \texttt{b}
will incur a page fault.  For \powm, we split the \texttt{mul\_mod} into $2$
pages and replicate the code for \texttt{set\_cond}.  Thus, call to from \powm
to \texttt{set\_cond} can be resolved to either of the pages.
% After applying this optimization to \powm, we reduce the
% overhead from $4000\times$ to $83\times$.
It is easy to see that since security guarantee of the compiler-transformed code holds true for 
the un-optimized program execution tree, it trivially holds true for the
reduced trees in the above two cases because O3A-B are replicating or merging the page access uniformly for all the inputs.  

\paragraph{O4: MUX Elimination.}
Our next optimization is based on the insight to eliminate the cost of 
the multiplexing operation itself by rearranging the code blocks. To achieve this, we
place the code blocks in the virtual pages to form an execution tree so that
all the transitions from one level to the other exhibit the same page fault.
This eliminates the multiplexing step altogether.  In the above example of blocks \texttt{a}, \texttt{b} and \texttt{c}, we place \texttt{a} and \texttt{b} into one page and \texttt{c} into another.  Thus, the control-flow 
from both \texttt{a} and \texttt{b} to \texttt{c} will  page fault in both the cases or none at all. We can chain together
such transitions for multiple levels in the tree,  such that all the blocks in
next level are always placed in a different pages. Figure~\ref{fig:powm-example} (c) 
shows the arrangement of functions in the  code staging area such
that the functions are grouped together in the same page. We
apply this to the execution sub-tree of \texttt{mul_mod}  function in \powm. 
% This helps us to reduces the overhead from $83\times$ to $3.8$\%
% (see Section~\ref{sec:eval}). 
% The security for this optimization is slightly different. 
% The main 

\subsection{Peephole Optimizations}

We apply a local peephole optimization to convert the control-dependent code to 
data-dependency which eliminates the need for code multiplexing. 

% . Several other peephole optimizations can also be borrowed to reduce
% the number of multiplexing operations in the transformation. We limit the discussion to  our case studies.
\begin{table*}
\caption{Summary of cryptographic implementations susceptible to \attacks, and their corresponding information leakage in gcc v4.8.2 and clang/llvm v3.4. $^{*}$ denotes that the leakage depends on the input. [$a:b$] denotes the split of S-Box where $a$ and $b$ is  percentage of table content across two different pages.
}

\centering
\resizebox{\textwidth}{!}{%

\begin{tabular}{|c|c|c|c|c|l|l|c|l|l|c|r|r|r|r|}
\hline
\multirow{2}{*}{{\bf Library}}                                                 & \multirow{2}{*}{{\bf Algo}} & \multirow{2}{*}{{\bf \begin{tabular}[c]{@{}c@{}}Secret\\ Entity\end{tabular}}}               & \multirow{2}{*}{{\bf \begin{tabular}[c]{@{}c@{}}Vulnerable \\ Routine\end{tabular}}} & \multicolumn{3}{c|}{\multirow{2}{*}{{\bf \begin{tabular}[c]{@{}c@{}}Vulnerable\\ Portion (gcc)\end{tabular}}}} & \multicolumn{3}{c|}{\multirow{2}{*}{{\bf \begin{tabular}[c]{@{}c@{}}Vulnerable\\ Portion (llvm)\end{tabular}}}} & \multirow{2}{*}{{\bf Input Bits}}                                        & \multicolumn{1}{c|}{\multirow{2}{*}{{\bf \begin{tabular}[c]{@{}c@{}}Leakage \\ (gcc)\end{tabular}}}} & \multicolumn{1}{c|}{\multirow{2}{*}{{\bf \%}}} & \multicolumn{1}{c|}{\multirow{2}{*}{{\bf \begin{tabular}[c]{@{}c@{}}Leakage\\ (llvm)\end{tabular}}}} & \multicolumn{1}{c|}{\multirow{2}{*}{{\bf \%}}} \\
                                                                               &                             &                                                                                              &                                                                                      & \multicolumn{3}{c|}{}                                                                                          & \multicolumn{3}{c|}{}                                                                                           &                                                                          & \multicolumn{1}{c|}{}                                                                                & \multicolumn{1}{c|}{}                          & \multicolumn{1}{c|}{}                                                                                & \multicolumn{1}{c|}{}                          \\ \hline
\multirow{16}{*}{\begin{tabular}[c]{@{}c@{}}Libgcrypt\\ (v1.6.3)\end{tabular}} & \multirow{2}{*}{AES}        & \multirow{4}{*}{Symmetric key}                                                               & \multirow{2}{*}{Encryption}                                                          & \multicolumn{3}{c|}{\multirow{2}{*}{2 T-Boxes {[}11:89{]}}}                                                    & \multicolumn{3}{c|}{\multirow{2}{*}{2 T-Boxes {[}50:50{]}}}                                                     & \multirow{2}{*}{\begin{tabular}[c]{@{}c@{}}128, 192,\\ 256\end{tabular}} & \multirow{2}{*}{25}                                                                                  & \multirow{2}{*}{14.01}                         & \multirow{2}{*}{8}                                                                                   & \multirow{2}{*}{4.51}                          \\
                                                                               &                             &                                                                                              &                                                                                      & \multicolumn{3}{c|}{}                                                                                          & \multicolumn{3}{c|}{}                                                                                           &                                                                          &                                                                                                      &                                                &                                                                                                      &                                                \\ \cline{2-2} \cline{4-15} 
                                                                               & \cast                       &                                                                                              & \multirow{2}{*}{Key Generation}                                                      & \multicolumn{3}{c|}{1 S-Box {[}38:62{]}}                                                                       & \multicolumn{3}{c|}{1 S-Box {[}48:52{]}}                                                                        & 128                                                                      & 3                                                                                                    & 2.34                                           & 2                                                                                                    & 1.56                                           \\ \cline{2-2} \cline{5-15} 
                                                                               & SEED                        &                                                                                              &                                                                                      & \multicolumn{3}{c|}{1 SS-Box {[}88:12{]}}                                                                      & \multicolumn{3}{c|}{1 SS-Box {[}27:73{]}}                                                                       & 128                                                                      & *6                                                                                                   & 4.69                                           & *4                                                                                                   & 3.13                                           \\ \cline{2-15} 
                                                                               & Stribog                     & \multirow{3}{*}{\begin{tabular}[c]{@{}c@{}}Password \\ used in \\ PBKDF2\end{tabular}}       & \multirow{3}{*}{Key Derivation}                                                      & \multicolumn{3}{c|}{4 S-Boxes {[}51:49{]}}                                                                     & \multicolumn{3}{c|}{4 S-Boxes {[}51:49{]}}                                                                      & 512                                                                      & 32                                                                                                   & 6.25                                           & 32                                                                                                   & 6.25                                           \\ \cline{2-2} \cline{5-15} 
                                                                               & Tiger                       &                                                                                              &                                                                                      & \multicolumn{3}{c|}{2 S-Boxes {[}53:47{]}}                                                                     & \multicolumn{3}{c|}{2 S-Boxes {[}58:42{]}}                                                                      & 512                                                                      & 4                                                                                                    & 0.78                                           & 4                                                                                                    & 0.78                                           \\ \cline{2-2} \cline{5-15} 
                                                                               & Whrilpool                   &                                                                                              &                                                                                      & \multicolumn{3}{c|}{4 S-Boxes {[}45:55{]}}                                                                     & \multicolumn{3}{c|}{4 S-Boxes {[}52:48{]}}                                                                      & 512                                                                      & 32                                                                                                   & 6.25                                           & 32                                                                                                   & 6.25                                           \\ \cline{2-15} 
                                                                               & \multirow{3}{*}{EdDSA}      & \multirow{3}{*}{\begin{tabular}[c]{@{}c@{}}Session key\\ (hence\\ Private key)\end{tabular}} & \multirow{3}{*}{Signing}                                                             & \multicolumn{3}{c|}{\multirow{3}{*}{ec\_mul}}                                                                  & \multicolumn{3}{c|}{\multirow{3}{*}{ec\_mul}}                                                                   & \multirow{3}{*}{512}                                                     & \multirow{3}{*}{512}                                                                                 & \multirow{3}{*}{100}                           & \multirow{3}{*}{512}                                                                                 & \multirow{3}{*}{100}                           \\
                                                                               &                             &                                                                                              &                                                                                      & \multicolumn{3}{c|}{}                                                                                          & \multicolumn{3}{c|}{}                                                                                           &                                                                          &                                                                                                      &                                                &                                                                                                      &                                                \\
                                                                               &                             &                                                                                              &                                                                                      & \multicolumn{3}{c|}{}                                                                                          & \multicolumn{3}{c|}{}                                                                                           &                                                                          &                                                                                                      &                                                &                                                                                                      &                                                \\ \cline{2-15} 
                                                                               & DSA                         & \multirow{2}{*}{Private key}                                                                 & Key generation                                                                       & \multicolumn{3}{c|}{\multirow{6}{*}{\powm}}                                                                    & \multicolumn{3}{c|}{\multirow{6}{*}{\powm}}                                                                     & 256                                                                      & *160                                                                                                 & 62.50                                          & *160                                                                                                 & 62.50                                          \\ \cline{2-2} \cline{4-4} \cline{11-15} 
                                                                               & Elgamal                     &                                                                                              & \multirow{5}{*}{\begin{tabular}[c]{@{}c@{}}Modular \\ exponentiation\end{tabular}}   & \multicolumn{3}{c|}{}                                                                                          & \multicolumn{3}{c|}{}                                                                                           & 400                                                                      & *238                                                                                                 & 59.50                                          & *238                                                                                                 & 59.50                                          \\ \cline{2-3} \cline{11-15} 
                                                                               & \multirow{4}{*}{RSA}        & \multirow{2}{*}{\begin{tabular}[c]{@{}c@{}}Private key\\ mod (p-1)\end{tabular}}             &                                                                                      & \multicolumn{3}{c|}{}                                                                                          & \multicolumn{3}{c|}{}                                                                                           & \multirow{2}{*}{2048}                                                    & \multirow{2}{*}{*1245}                                                                               & \multirow{2}{*}{60.79}                         & \multirow{2}{*}{*1245}                                                                               & \multirow{2}{*}{60.79}                         \\
                                                                               &                             &                                                                                              &                                                                                      & \multicolumn{3}{c|}{}                                                                                          & \multicolumn{3}{c|}{}                                                                                           &                                                                          &                                                                                                      &                                                &                                                                                                      &                                                \\ \cline{3-3} \cline{11-15} 
                                                                               &                             & \multirow{2}{*}{\begin{tabular}[c]{@{}c@{}}Private key\\ mod (q-1)\end{tabular}}             &                                                                                      & \multicolumn{3}{c|}{}                                                                                          & \multicolumn{3}{c|}{}                                                                                           & \multirow{2}{*}{2048}                                                    & \multirow{2}{*}{*1247}                                                                               & \multirow{2}{*}{60.89}                         & \multirow{2}{*}{*1247}                                                                               & \multirow{2}{*}{60.89}                         \\
                                                                               &                             &                                                                                              &                                                                                      & \multicolumn{3}{c|}{}                                                                                          & \multicolumn{3}{c|}{}                                                                                           &                                                                          &                                                                                                      &                                                &                                                                                                      &                                                \\ \hline
\multirow{2}{*}{\begin{tabular}[c]{@{}c@{}}OpenSSL\\ (v1.0.2)\end{tabular}}    & \cast                       & \multirow{2}{*}{Symmetric key}                                                               & \multirow{2}{*}{Key generation}                                                      & \multicolumn{3}{c|}{1 S-Box {[}55:45{]}}                                                                       & \multicolumn{3}{c|}{1 S-Box {[}84:16{]}}                                                                        & 128                                                                      & 2                                                                                                    & 1.56                                           & *6                                                                                                   & 4.69                                           \\ \cline{2-2} \cline{5-15} 
                                                                               & SEED                        &                                                                                              &                                                                                      & \multicolumn{3}{c|}{1 SS-Box {[}47:53{]}}                                                                      & \multicolumn{3}{c|}{1 SS-Box {[}67:33{]}}                                                                       & 128                                                                      & 16                                                                                                   & 12.50                                          & *6                                                                                                   & 4.69                                           \\ \hline
\multicolumn{12}{|r|}{{\bf Average}}                                                                                                                                                                                                                                                                                                                                                                                                                                                                                                                                                                                                                                                                                    & {\bf 28.02}                                    &                                                                                                      & {\bf 25.64}                                    \\ \hline
\end{tabular}
% \vspace{-.3cm}
}
\label{tab:attacks}
\vspace{-.6cm}
\end{table*}

% \vspace{-.4cm}
\begin{figure}[t]
\centering
\begin{minipage}[r]{0.165\textwidth}

\begin{lstlisting}
if (c) {
  result = result*2;
}
\end{lstlisting}%
\centering (a)
\end{minipage}%
% \hfil
\begin{minipage}[c]{0.1\textwidth}
\centering $\implies$
% \fbox{\parbox[m]{.8\textwidth}{\centering $\implies$}}
\end{minipage}%
% \hfil
\begin{minipage}[l]{0.2\textwidth}
\begin{lstlisting}
staging_area[0] = result; 
staging_area[1] = result*2; 
result = staging_area[c];
\end{lstlisting}%
\centering (b)
\end{minipage}%
\vspace{-.3cm}
\caption{Example for O5: Control-to-Data Dependency Transformation.}
\label{lst1}
\vspace{-.2cm}
\end{figure}

\paragraph{O5: Control-to-Data Dependency Transformation.}
Masking data page accesses is easier and hence we can convert the input
dependent code accesses to data accesses. For example, the if-condition on
value of \texttt{c} in Figure~\ref{lst1}~(a) can be rewritten as
Figure~\ref{lst1}~(b). Specifically, we perform an if-conversion such that
the code is always executed and the condition is used to decide whether to
retain the results or discard them~\cite{x86-timing}. In the case of \eddsa,
we first fetch the value of \texttt{res} into $SA_{data}$ (Refer to
Figure~\ref{fig:egcf} for code details).  We execute \texttt{add_points}
unconditionally and  we use \texttt{test_bit} as a selector to decide if the
value in $SA_{data}$ is to be used. In the case where \texttt{test_bit}
returns true, the actual \texttt{res} in $SA_{data}$ is used in the operation
and is updated, else it is discarded. The page fault pattern will be
deterministic since \texttt{add_points} will be executed on all iterations of
the loop and the operand of the function is always from $SA_{data}$. This
optimization is applied before the compiler transformation, hence its
security follows from the basic security invariant outlined in
Section~\ref{sec:automation}.

All our strategies \textbf{O1}-\textbf{O5} are supported by our compiler augmentation 
with programmer directives. Note that, our 
optimization strategies are sound --- the compiler still asserts that the
transformation preserves the \property of the program.  We discuss 
the empirical effectiveness of these strategies in Section~\ref{sec:opteval}.
\section{Evaluation}
\label{sec:eval}

% \usepackage{multirow}
\begin{table*}[t]
\caption{Performance Summary. Columns 3, 5, 12 denotes the number of page faults incurred at runtime. Columns 10 and 14 represent the total percentage overhead. $>$ symbol denotes the program did not complete within $10$ hours after which we terminated it. A negative overhead means patched code executes faster than the baseline. Tc and Te denote the time spent in preparing the staging area and actual execution respectively.}
\label{tab:performance}
\centering
\resizebox{\textwidth}{!}{%
\begin{tabular}{|c|c|l|r|c|r|r|r|r|r|l|c|r|r|}
\hline
\multicolumn{1}{|l|}{\multirow{3}{*}{{\bf Library}}}                          & \multirow{3}{*}{{\bf Cases}} & \multicolumn{2}{c|}{{\bf Vanilla}}                                                                  & \multicolumn{6}{c|}{{\bf Deterministic Multiplexing}}                                                                                                                                                                                                                                                       & \multicolumn{4}{c|}{{\bf 

\begin{tabular}[c]{@{}c@{}}Optimized \\ Deterministic Multiplexing\end{tabular}}}                                                         \\ \cline{3-14} 
\multicolumn{1}{|l|}{}                                                        &                              & \multicolumn{1}{c|}{\multirow{2}{*}{{\bf PF}}} & \multicolumn{1}{c|}{\multirow{2}{*}{{\bf T (ms)}}} & \multirow{2}{*}{{\bf PF}} & \multicolumn{1}{l|}{\multirow{2}{*}{{\bf Tc (ms)}}} & \multicolumn{1}{l|}{\multirow{2}{*}{{\bf Te (ms)}}} & \multicolumn{1}{c|}{\multirow{2}{*}{{\bf T (ms)}}} & \multicolumn{1}{l|}{\multirow{2}{*}{{\bf Tc / T (\%)}}} & \multicolumn{1}{c|}{\multirow{2}{*}{{\bf Ovh (\%)}}} & \multirow{2}{*}{{\bf Opt}} & \multirow{2}{*}{{\bf PF}} & \multicolumn{1}{c|}{\multirow{2}{*}{{\bf T (ms)}}} & \multicolumn{1}{c|}{\multirow{2}{*}{{\bf Ovh (\%)}}} \\
\multicolumn{1}{|l|}{}                                                        &                              & \multicolumn{1}{c|}{}                          & \multicolumn{1}{c|}{}                              &                           & \multicolumn{1}{l|}{}                               & \multicolumn{1}{l|}{}                               & \multicolumn{1}{c|}{}                              & \multicolumn{1}{l|}{}                                   & \multicolumn{1}{c|}{}                                &                            &                           & \multicolumn{1}{c|}{}                              & \multicolumn{1}{c|}{}                                \\ \hline
\multirow{9}{*}{\begin{tabular}[c]{@{}c@{}}Libgcrypt\\ (v1.6.3)\end{tabular}} & AES                          & 4 - 5                                          & 4.711                                              & 4                         & 7.357                                               & 4.013                                               & 11.370                                             & 64.70                                                   & {\bf 141.35}                                         & O1,O2                      & 4                         & 4.566                                              & {\bf -3.08}                                          \\ \cline{2-14} 
                                                                              & \cast                        & 2                                              & 3.435                                              & 2                         & 8.050                                               & 2.578                                               & 10.629                                             & 75.74                                                   & {\bf 209.47}                                         & O1,O2                      & 1                         & 3.086                                              & {\bf -10.15}                                         \\ \cline{2-14} 
                                                                              & EdDSA                        & 0                                              & 10498.674                                          & 0                         & \multicolumn{2}{c|}{\multirow{3}{*}{---}}                                                                 & \multirow{3}{*}{\textgreater 10 hrs}               & \multicolumn{1}{c|}{\multirow{3}{*}{---}}               & {\bf \textgreater 300000}                            & O5                         & 0                         & 13566.122                                          & {\bf 29.22}                                          \\ \cline{2-5} \cline{10-14} 
                                                                              & \multirow{2}{*}{\powm}       & \multirow{2}{*}{0}                             & \multirow{2}{*}{5318.501}                          & \multirow{2}{*}{0}        & \multicolumn{2}{c|}{}                                                                                     &                                                    & \multicolumn{1}{c|}{}                                   & \multirow{2}{*}{{\bf \textgreater 400000}}           & O3                         & 0                         & 399614.244                                         & 7413.66                                              \\ \cline{11-14} 
                                                                              &                              &                                                &                                                    &                           & \multicolumn{2}{c|}{}                                                                                     &                                                    & \multicolumn{1}{c|}{}                                   &                                                      & O4                         & 0                         & 5513.712                                           & {\bf 3.67}                                           \\ \cline{2-14} 
                                                                              & SEED                         & 2                                              & 1.377                                              & 2                         & 4.559                                               & 1.057                                               & 5.615                                              & 81.18                                                   & {\bf 307.79}                                         & O1, O2                     & 1                         & 1.311                                              & {\bf -4.80}                                          \\ \cline{2-14} 
                                                                              & Stribog                      & 5                                              & 27.397                                             & 5                         & 329.743                                             & 10.836                                              & 340.579                                            & 96.82                                                   & {\bf 1143.13}                                        & O1, O2                     & 4                         & 28.563                                             & {\bf 4.26}                                           \\ \cline{2-14} 
                                                                              & Tiger                        & 3                                              & 2.020                                              & 3                         & 64.482                                              & 0.546                                               & 65.029                                             & 99.16                                                   & {\bf 3119.69}                                        & O1, O2                     & 2                         & 1.840                                              & {\bf -8.89}                                          \\ \cline{2-14} 
                                                                              & Whirlpool                    & 5                                              & 27.052                                             & 5                         & 141.829                                             & 10.174                                              & 151.490                                            & 93.28                                                   & {\bf 459.99}                                         & O1, O2                     & 4                         & 23.744                                             & {\bf -12.23}                                         \\ \hline
\multirow{2}{*}{\begin{tabular}[c]{@{}c@{}}OpenSSL\\ (v1.0.2)\end{tabular}}   & \cast                        & 2                                              & 1.147                                              & 2                         & 0.815                                               & 0.690                                               & 1.505                                              & 54.139                                                  & {\bf 31.19}                                          & O1, O2                     & 1                         & 0.880                                              & {\bf -23.28}                                         \\ \cline{2-14} 
                                                                              & SEED                         & 2                                              & 0.651                                              & 2                         & 0.511                                               & 0.576                                               & 1.087                                              & 47.024                                                  & {\bf 67.10}                                          & O1, O2                     & 1                         & 0.639                                              & {\bf -1.74}                                          \\ \hline
\multicolumn{9}{|r|}{{\bf Average Performance Overhead}}                                                                                                                                                                                                                                                                                                                                                                                                                  & {\bf 70547.971}                                      & \multicolumn{3}{l|}{}                                                                                       & {\bf -2.70}                                          \\ \hline
\end{tabular}
}
% \vspace{-.3cm}
\vspace{-.5cm}
\end{table*}

\paragraph{Evaluation Goals.}
We aim to evaluate the effectiveness of our proposed solutions for following
main goals:
\vspace{0.2cm}
\begin{itemize}
\squish
	\item{Does our defense apply to all of our case studies?}
	\item{What are the performance trade-offs of our defense?}
	\item{How much performance improvements do developer-assisted transformation offer?}
\end{itemize}

\paragraph{Platform.}
SGX hardware is not yet fully rolled out and is not publicly available for
experimentation. As a recourse, we conduct all our experiments on
\podarch~\cite{podarch}; a system similar to previous hypervisor
solutions~\cite{overshadow} and conceptually similar to SGX.  Our machine is a
Dell Latitude 6430u host, configured with Intel(R) Core(TM) i7-3687U 2.10GHz
CPU, 8GB RAM. We configure \podarch with one CPU, 2GB RAM and 64-bit Linux
3.2.53 Kernel on Debian Jessie for all the experiments. We use LLVM v3.4 with
the default optimization flags for compiling our vanilla and patched case
studies. All the results are averaged over five runs. 
% For consistency, we set the native CPU scaling to performance mode.

\subsection{Case Studies}
\label{sec:cases}

\paragraph{Selection Criteria.}
Our defense techniques can be applied to an application if it satisfies the
conditions of balanced-execution tree. We checked the programs FreeType,
Hunspell, and libjpeg discussed in~\cite{cca-sgx}, they exhibit unbalanced
execution tree.  Transforming these programs to exhibit balanced execution
tree causes an unacceptable loss in the performance, even without our
defense~\cite{oblv-shruti}. Hence, we limit our evaluation to cryptographic
implementations.

We present our results from the study of a general purpose cryptographic
library \libgcrypt v1.6.3 which is used in GnuPG and a SSL implementation
library \openssl v1.0.2~\cite{libgcrypt, gnupg, openssl}. We analyzed the
programs  compiled with the two most-used compiler toolchains: \texttt{gcc}
v4.8.2 and LLVM v3.4. For both the compilers, we statically compiled all our
programs with the default optimization and security flags specified in their
\texttt{Makefile}. Of the $24$ routines we analyze in total from both the
libraries, $10$ routines are vulnerable to \attacks on both the compilers.
Since our emphasis is not on the attacks, we highlight the important findings
below. Table~\ref{tab:attacks}  summarizes the results of our study.
Interested readers can refer to Appendix~\ref{appx:attacks} for the
experimental details of each case study attack.

% \begin{itemize}
% 	\squish
%     \item{\em Asymmetric-Key Crypto Implementations.} ECDSA, \eddsa, ElGamal, Diffie-Hellman, DSA, RSA.
% 	\item{\em Symmetric-key Crypto Implementations.} AES, Blowfish, Camellia, \cast, DES, 3DES, IDEA, RC5, \seed, Serpent, Twofish. 
% 	\item{\em Hash Function Implementations.} SHA512, Stribog-512, Tiger, Whirlpool.
% \end{itemize}

\vspace{0.2cm}
\begin{itemize}
\squish
	\item {\bf No Leakage}.  
	In \libgcrypt implementations of Blowfish, Camellia, DES, 3DES, IDEA, RC5,
	Serpent, Twofish, ECDSA, and SHA512, all the input-dependent 
	code and data memory accesses are confined within a page for the
	sensitive portions. Similarly AES, Blowfish, Camellia, DES, 3DES, IDEA,
	RC5, Serpent, Twofish, DSA, RSA, and SHA512 in \openssl do not exhibit
	leakage via page fault side channel.

	\item {\bf Leakage via input dependent code page access}. 
	In \libgcrypt, \eddsa and \powm exhibit input dependent code access across
	pages and are vulnerable to \attacks. The \powm function is used in ElGamal,
	DSA and RSA which leaks bits of information about the secret exponents.

	\item {\bf Leakage via input dependent data page access}.  
	In case of AES, \cast, \seed, Stribog, Tiger and Whirlpool implementations
	in \libgcrypt, at least one of the S-Boxes crosses page boundary and leaks
	information about the secret inputs.  Similarly, implementations of \cast
	and \seed in \openssl are also vulnerable.

\end{itemize}

\subsection{Application to Case Studies}
We transform the $8$ \libgcrypt and $2$ \openssl vulnerable implementations of
our case studies in our evaluation.

\paragraph{Compiler Toolchain Implementation.}
We implement our automation tool in LLVM 3.4 and Clang 3.4 C / C++ front-end
to transform C / C++ applications~\cite{llvm, clang}. For our case studies, we
log all the analysis information which is used for the layout analysis and
also to facilitate our developer-assisted improvements study. Our
transformation pass applies \interpreting to the programs at the LLVM IR
level.  Table~\ref{tab:stats} shows the number of functions, execution 
blocks, loops, variables and total size of code and data
staging area.% as required by our toolchain.

%  Of these $24$ programs, details of
% the $8$ vulnerable cryptographic programs is outlined in Table~\ref{tab:attacks}.

% \usepackage{multirow}
\begin{table}[]
\caption{Analysis Summary. Column 3, 4, 5, 6, 7 denote the total number of functions, execution blocks, loops, variables and lines of code in the sensitive area respectively. %Column 4 denotes the average size (in number of instruction count) of the execution blocks. 
Column 8 gives the number pages allocated for staging area instances. Column 9, 10 gives the total number of function calls and accesses to the staging area at runtime respectively.}
\label{tab:stats}
\centering
\tiny

\begin{tabular}{|l|l|r|r|r|r|r|r|r|r|}
\hline
\multirow{4}{*}{\textbf{Lib}}                                         & \multicolumn{1}{c|}{\multirow{4}{*}{\textbf{Cases}}} & \multicolumn{1}{c|}{\multirow{4}{*}{\textbf{\begin{tabular}[c]{@{}c@{}}\#\\ F\end{tabular}}}} & \multicolumn{1}{c|}{\multirow{4}{*}{\textbf{\begin{tabular}[c]{@{}c@{}}\#\\ EB\end{tabular}}}} & \multicolumn{1}{c|}{\multirow{4}{*}{\textbf{\begin{tabular}[c]{@{}c@{}}\#\\ L\end{tabular}}}} & \multicolumn{1}{c|}{\multirow{4}{*}{\textbf{\begin{tabular}[c]{@{}c@{}}\#\\ V\end{tabular}}}} & \multicolumn{1}{c|}{\multirow{4}{*}{\textbf{LoC}}} & \multicolumn{1}{c|}{\multirow{4}{*}{\textbf{\begin{tabular}[c]{@{}c@{}}\# \\ P\end{tabular}}}} & \multicolumn{1}{c|}{\multirow{4}{*}{\textbf{\begin{tabular}[c]{@{}c@{}}\#\\ RTF\end{tabular}}}} & \multicolumn{1}{c|}{\multirow{4}{*}{\textbf{\begin{tabular}[c]{@{}c@{}}\# \\ MUX\end{tabular}}}} \\
                                                                      & \multicolumn{1}{c|}{}                                & \multicolumn{1}{c|}{}                                                                         & \multicolumn{1}{c|}{}                                                                          & \multicolumn{1}{c|}{}                                                                         & \multicolumn{1}{c|}{}                                                                         & \multicolumn{1}{c|}{}                              & \multicolumn{1}{c|}{}                                                                          & \multicolumn{1}{c|}{}                                                                           & \multicolumn{1}{c|}{}                                                                            \\
                                                                      & \multicolumn{1}{c|}{}                                & \multicolumn{1}{c|}{}                                                                         & \multicolumn{1}{c|}{}                                                                          & \multicolumn{1}{c|}{}                                                                         & \multicolumn{1}{c|}{}                                                                         & \multicolumn{1}{c|}{}                              & \multicolumn{1}{c|}{}                                                                          & \multicolumn{1}{c|}{}                                                                           & \multicolumn{1}{c|}{}                                                                            \\
                                                                      & \multicolumn{1}{c|}{}                                & \multicolumn{1}{c|}{}                                                                         & \multicolumn{1}{c|}{}                                                                          & \multicolumn{1}{c|}{}                                                                         & \multicolumn{1}{c|}{}                                                                         & \multicolumn{1}{c|}{}                              & \multicolumn{1}{c|}{}                                                                          & \multicolumn{1}{c|}{}                                                                           & \multicolumn{1}{c|}{}                                                                            \\ \hline
\multirow{8}{*}{\begin{tabular}[c]{@{}l@{}}Libg\\ crypt\end{tabular}} & AES                                                  & 1                                                                                             & 1                                                                                              & 0                                                                                             & 22                                                                                            & 272                                                & 3                                                                                              & 1                                                                                               & 112                                                                                              \\ \cline{2-10} 
                                                                      & \cast                                                & 1                                                                                             & 1                                                                                              & 0                                                                                             & 11                                                                                            & 47                                                 & 3                                                                                              & 2                                                                                               & 320                                                                                              \\ \cline{2-10} 
                                                                      & EdDSA                                                & 31                                                                                            & 701                                                                                            & 56                                                                                            & 178                                                                                           & 1212                                               & 2                                                                                              & 1                                                                                               & 60725                                                                                            \\ \cline{2-10} 
                                                                      & \powm                                                & 20                                                                                            & 297                                                                                            & 47                                                                                            & 126                                                                                           & 796                                                & 2                                                                                              & 1                                                                                               & 57660                                                                                            \\ \cline{2-10} 
                                                                      & SEED                                                 & 1                                                                                             & 11                                                                                             & 1                                                                                             & 17                                                                                            & 56                                                 & 3                                                                                              & 1                                                                                               & 128                                                                                              \\ \cline{2-10} 
                                                                      & Stribog                                              & 1                                                                                             & 1                                                                                              & 0                                                                                             & 5                                                                                             & 31                                                 & 5                                                                                              & 250                                                                                             & 2000                                                                                             \\ \cline{2-10} 
                                                                      & Tiger                                                & 1                                                                                             & 1                                                                                              & 0                                                                                             & 12                                                                                            & 18                                                 & 3                                                                                              & 312                                                                                             & 2496                                                                                             \\ \cline{2-10} 
                                                                      & Whirlpool                                            & 1                                                                                             & 3                                                                                              & 1                                                                                             & 16                                                                                            & 112                                                & 6                                                                                              & 6                                                                                               & 7680                                                                                             \\ \hline
\multirow{2}{*}{\begin{tabular}[c]{@{}l@{}}Open\\ SSL\end{tabular}}   & \cast                                                & 1                                                                                             & 12                                                                                             & 4                                                                                             & 14                                                                                            & 93                                                 & 3                                                                                              & 1                                                                                               & 160                                                                                              \\ \cline{2-10} 
                                                                      & SEED                                                 & 1                                                                                             & 1                                                                                              & 0                                                                                             & 13                                                                                            & 58                                                 & 3                                                                                              & 1                                                                                               & 128                                                                                              \\ \hline
\end{tabular}
\end{table}

\paragraph{Empirical Validation.}
Our applications are compiled into static binaries for testing. We run these
executables on \podarch~\cite{podarch} which is implemented on QEMU emulator,
and only supports static linking. To test our patched applications, we execute
the standard regression test-suite available with the cryptographic libraries
(\texttt{make check}). To empirically validate that our defenses work, we
ensure that the page fault profile of patched executions  under all test
inputs is indistinguishable w.r.t. page access profiles. To verify the
correctness, we analyze the page fault access patterns in the transformed
application using a PinTool~\cite{pintool} that logs all instructions and
memory accesses. We have analyzed the PinTools logs and report that our
\interpreting produces indistinguishable page access profiles for all
regression and test inputs.

\subsection{Performance Evaluation}

\paragraph{Normalized Baseline.}
To ensure that the choice of our evaluation platform (\podarch) does not significantly 
bias the overheads, we conduct two sets of measurements. First, we run the
unmodified \openssl and \libgcrypt implementations on \podarch and measure the execution
time. This forms the baseline for all our performance measurements.  Column
$3$, $4$ in Table~\ref{tab:performance}  shows the number of page faults and
the execution time for vanilla code in \podarch. Second, to check that the
overheads of our defenses are not an artifact of \podarch, we also run our
vanilla and modified binaries  on native Intel CPU Intel Core i7-2600 CPU. The
overheads on a native CPU  are similar to that on \podarch and deviate only
within a range of $1\%$. This confirms that our baseline of \podarch is
unbiased and does not skew our experimental results.

\paragraph{Overhead.}
We calculate the overhead by comparing the baseline performance of unmodified
code against the execution time of the patched application functions.   We use
input patterns to represent the best, worst and average case executions of the
application, specifically, inputs with (a) all 0s, (b) all 1s, (c) random
number of 0s and 1s, and (d) all the regression tests from the built-in test-
suite.

% (a) 
% \begin{itemize}
% \squish
% 	\item {Zeros:} The input consists of all 0s.% except for the two most-significant bits that are set to one. 
% 	\item {Ones:} The input consists of all 1s. 
% 	\item {Random:} The input consists of random number of 1s and 0s.
% 	\item {All the regression tests:} Inputs from in the library test suite. 
% \end{itemize}

% \paragraph{\Interpreting.}
% We manually patch all the $8$ case studies discussed in
% Section~\ref{sec:cases} to 

The applications patched with the \interpreting technique incurs an average
overhead of $705\times$ and up to maximum overhead of $4000\times$ in case of
\powm (Column $10$ in Table~\ref{tab:performance}).  To investigate the main
sources of these overheads we measure the break-down for the fetch step and the
execute step in \interpreting. We observe that the overhead is mainly
dominated by the copying of data to and from the staging area in the fetch step
(Column $6$ and $9$ in Table~\ref{tab:performance}), and accounts for $76.5\%$
out of the total overhead on average. We notice that the fetch step time is
especially high for cases like Stribog and Tiger where it accounts for
$96.82\%$ and $99.16\%$ of the overhead.

\subsection{Effectiveness of Optimizations}
%Developer-assisted Transformations}
\label{sec:opteval}
We apply the developer-assisted strategies discussed in
Section~\ref{sec:optimization} to experimentally validate and demonstrate
their effectiveness. They reduce the  average overhead from $705\times$ to
$-2.7\%$ for  our $10$ case studies; $29.22\%$ in the worst case. In the case of \powm, \textbf{O3} reduces
the performance overhead from $4000\times$ to $74\times$.  With \textbf{O4} we
completely remove memory copying for code determinization which reduces  the
overhead from $74\times$ to $3.67$\%. We apply  \textbf{O1} to the $8$ cases
of input dependent data page access to reduce the number of copy operations.
Further we also apply \textbf{O2} to reorder the lookup table layout, such
that after the developer-assisted transformations are in place, the execution
incurs lower page faults. In fact, our patched version executes faster than
the baseline code (as denoted by negative overhead in Column $14$ in
Table~\ref{tab:performance}) for $7$ cases.  After manual inspection, this is
explained because in the patched code, the lookup tables take up less number of pages 
which reduces the total number of page faults incurred during the
execution (Column $12$ in Table~\ref{tab:performance}).  On the other hand, in
the vanilla case, the program incurs more page faults which is a
costly operation. Thus, eliminating this cost results in a negative overhead.
For \eddsa, we directly apply the peephole optimization \textbf{O5} which
transforms the input dependent code access to data access. This reduces the
overhead from $3000\times$  to $29.22\%$.

% \paragraph{\Bucketing.}
% As compared to \interpreting, \bucketing incurs a lower overhead  of on an
% $6.77$\% on an average and $27.25$\% in the worst case. (See column $4$ and
% $7$ in Table~\ref{tab:performance}). The overhead incurred is mainly due to
% the extra pages that are loaded at the beginning  of the sensitive code
% section.
% \section{Alternative Defenses}
\section{Hardware-enabled Defenses}
\label{sec:discussion}

So far we have discussed purely software solutions. Readers might wonder if
\attacks can be mitigated with hardware support. Here, we briefly discuss  an
alternative hardware-assisted defense which guarantees \execution at a 
worst-case cost of $6.77$\% on our benchmarks.

% \begin{figure*}[t]
% \centering
% \begin{minipage}[t]{0.32\textwidth}
% \vspace{0pt}
% \includegraphics[width=6cm,height=3cm]{figures/new/7}
% \end{minipage}
% \hfill
% \begin{minipage}[t]{0.62\textwidth}
% \vspace{0pt}
% \tiny
% \input{tables/security}
% \end{minipage} 
% \caption{(a) \Bucketing Design. 
% (1) Enclave registers a bucket for $P_1, P_2, P_3, P_4$ . 
% (2) CPU directly reports the fault within a bucket to the \vdf. 
% (3) \vdf fakes access for time \texttt{t - k} and sends command to terminate.
% (4) enclave fault handler terminates the enclave.
% (b) Evaluation. Column 2 denotes the Bucket size (Code + Data). Columns 5 and 7 denote average execution time and deviation in benign OS. Columns 8-10 denote total time spent (before inducing page fault + in \vdf) for 3 scenarios.}
% \vspace{-15pt}
% \label{fig:bucketing}
% \end{figure*} 
% \usepackage{multirow}
\begin{table*}[t]
\caption{Evaluation. Column 2 denotes the bucket size (Code + Data). Columns 5 and 7 denote average execution time and deviation in benign OS. Columns 8-10 denote total time spent for 3 test-case scenarios that stress the corner cases in \libgcrypt.  Both the  executions exhibit no statistically significant differences.}% This confirms that that our termination strategy is safe.}
% \caption{\Bucketing.
% Column 2 denotes the Bucket size (Code + Data). Columns 5 and 7 denote average execution time and deviation in benign OS. Columns 8-10 denote total time spent (before inducing page fault + in \vdf) for 3 scenarios.}
\centering
% % \tiny
\label{tab:sec-perf}
\resizebox{\textwidth}{!}{%

\begin{tabular}{|l|c|l|c|r|r|r|r|r|r|r|}
\hline
\multicolumn{1}{|c|}{\multirow{4}{*}{{\bf Cases}}} & \multicolumn{2}{c|}{\multirow{4}{*}{{\bf \begin{tabular}[c]{@{}c@{}}Bucket\\ Size\end{tabular}}}} & \multirow{4}{*}{{\bf \begin{tabular}[c]{@{}c@{}}PF\\ Handler\\ (Bytes)\end{tabular}}} & \multicolumn{4}{c|}{\multirow{2}{*}{{\bf Benign OS}}}                                                                                                                                                                                                                                                                               & \multicolumn{3}{c|}{\multirow{2}{*}{{\bf Malicious OS}}}                                                                                                        \\
\multicolumn{1}{|c|}{}                             & \multicolumn{2}{c|}{}                                                                             &                                                                                       & \multicolumn{4}{c|}{}                                                                                                                                                                                                                                                                                                               & \multicolumn{3}{c|}{}                                                                                                                                           \\ \cline{5-11} 
\multicolumn{1}{|c|}{}                             & \multicolumn{2}{c|}{}                                                                             &                                                                                       & \multicolumn{1}{c|}{\multirow{2}{*}{{\bf \begin{tabular}[c]{@{}c@{}}Vanilla\\ Time (ms)\end{tabular}}}} & \multicolumn{1}{c|}{\multirow{2}{*}{{\bf \begin{tabular}[c]{@{}c@{}}Contractual\\ Time (ms)\end{tabular}}}} & \multicolumn{1}{c|}{\multirow{2}{*}{{\bf Ovh (\%)}}} & \multicolumn{1}{c|}{\multirow{2}{*}{{\bf Dev (\%)}}} & \multicolumn{1}{c|}{\multirow{2}{*}{{\bf T1 (ms)}}} & \multicolumn{1}{c|}{\multirow{2}{*}{{\bf T2 (ms)}}} & \multicolumn{1}{c|}{\multirow{2}{*}{{\bf T3 (ms)}}} \\
\multicolumn{1}{|c|}{}                             & \multicolumn{2}{c|}{}                                                                             &                                                                                       & \multicolumn{1}{c|}{}                                                                                   & \multicolumn{1}{c|}{}                                                                                       & \multicolumn{1}{c|}{}                                & \multicolumn{1}{c|}{}                                & \multicolumn{1}{c|}{}                               & \multicolumn{1}{c|}{}                               & \multicolumn{1}{c|}{}                               \\ \hline
AES                                                & \multicolumn{2}{c|}{3 + 3}                                                                        & 274                                                                                   & 4.157                                                                                                   & 4.161                                                                                                       & {\bf 0.107}                                          & 4.689                                                & 4.287                                               & 4.179                                               & 4.059                                               \\ \hline
\cast                                              & \multicolumn{2}{c|}{1 + 2}                                                                        & 231                                                                                   & 2.901                                                                                                   & 2.969                                                                                                       & {\bf 2.34}                                           & 9.938                                                & 3.054                                               & 3.003                                               & 2.845                                               \\ \hline
EdDSA                                              & \multicolumn{2}{c|}{19 + 1}                                                                       & 204                                                                                   & 9729.526                                                                                                & 9754.806                                                                                                    & {\bf 0.260}                                          & 35.952                                               & 9960.311                                            & 9815.837                                            & 10146.534                                           \\ \hline
\powm                                              & \multicolumn{2}{c|}{21 + 1}                                                                       & 256                                                                                   & 4783.997                                                                                                & 4813.028                                                                                                    & {\bf 0.607}                                          & 12.225                                               & 5155.958                                            & 5103.789                                            & 5224.345                                            \\ \hline
SEED                                               & \multicolumn{2}{c|}{2 + 2}                                                                        & 261                                                                                   & 1.269                                                                                                   & 1.381                                                                                                       & {\bf 8.917}                                          & 4.821                                                & 1.337                                               & 1.392                                               & 1.333                                               \\ \hline
Stribog                                            & \multicolumn{2}{c|}{1 + 5}                                                                        & 253                                                                                   & 0.803                                                                                                   & 0.874                                                                                                       & {\bf 8.957}                                          & 1.940                                                & 0.863                                               & 0.879                                               & 0.887                                               \\ \hline
Tiger                                              & \multicolumn{2}{c|}{1 + 3}                                                                        & 244                                                                                   & 0.506                                                                                                   & 0.644                                                                                                       & {\bf 27.255}                                         & 4.876                                                & 0.667                                               & 0.659                                               & 0.675                                               \\ \hline
Whirlpool                                          & \multicolumn{2}{c|}{1 + 5}                                                                        & 245                                                                                   & 12.680                                                                                                  & 13.409                                                                                                      & {\bf 5.746}                                          & 1.338                                                & 13.559                                              & 13.451                                              & 13.308                                              \\ \hline
\multicolumn{6}{|r|}{{\bf Average}}                                                                                                                                                                                                                                                                                                                                                                                                                                    & {\bf 6.77}                                           & \multicolumn{4}{l|}{}                                                                                                                                                                                                  \\ \hline
\end{tabular}
}
\vspace{-.5cm}
\end{table*}

\subsection{Our Proposal: \Bucketing}
We propose a hardware-software technique wherein the enclave is guaranteed by
the hardware that certain virtual addresses will always be mapped to physical
memory during its execution. The enclave application is coded optimistically
assuming that the OS will always allocate specific number of physical pages to
it while executing its sensitive code blocks. The enclave informs its memory
requirements to the OS via a callback mechanism.  These requirements act as a
{\em contract} if the OS agrees, or else the OS can refuse to start execution
of the enclave. The enclave states the set of virtual addresses explicitly to
the OS before starting its sensitive computation. The CPU acts as a contract
mediator and is responsible for enforcing this guarantee on the OS. We term
such an execution as {\em contractual execution}. Note that  the contract is
not a hard guarantee i.e., the enclave cannot pin the pages in physical memory
to launch a denial-of-service attack on the OS. In fact, the OS has the
flexibility to take back pages as per its own scheduling policy. However, when
the CPU observes that OS has deviated from the contract --- either genuinely
or by injecting random faults,  it immediately reports the contract  violation
to the enclave. This needs two types of changes in the hardware (a) support
for notifying the enclave about its own page faults and  (b) guaranteeing a
safe mechanism for enclave to mitigate the contract violation.

\paragraph{Contract Enforcement in SGX.}
In a traditional CPU as well as in original SGX specification~\cite{sgx1}, all
page faults are reported directly to the OS without the intervention of the
faulting process. Thus, the process is unaware of its own page faults. This
makes it impossible for the enclave to detect \attacks.  For \bucketing, the
hardware needs to report its faults to the process instead, which calls for a
change in the page fault semantics. A limited amount of support is already
available for this in SGX. As per the new amendments in Revision 2, SGX can
now notify an enclave about its page faults by setting the
\texttt{SECS.MISCSELECT.EXINFO} bit~\cite{sgx2, haven}. When an enclave
faults, the SGX hardware notifies the enclave about the fault, along with the
virtual address, type of fault, the permissions of the page, register context.
We can think of implementing contractual execution on SGX directly by setting
the SGX configuration bit such that when there is a page fault, the enclave
will be notified directly by the CPU.  The benign OS is expected to respect
the contract and never swap out the pages during the execution. However a
malicious OS may swap out pages, in which case the  CPU is responsible for
reporting page faults for these pages to the enclave directly.

\begin{figure}
\centering
\epsfig{file=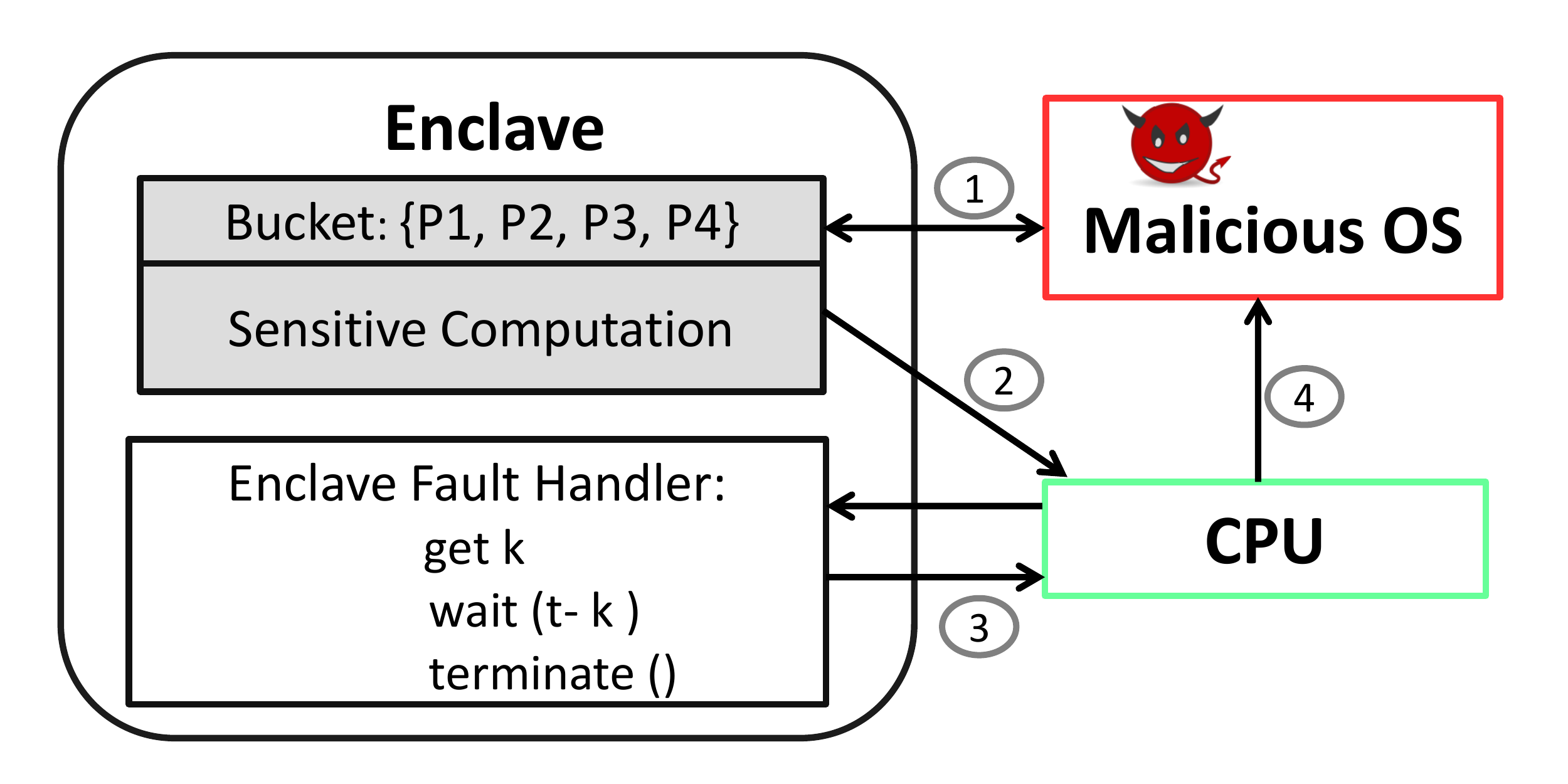, width=7.5cm, height=3.7cm} %scale=0.2]
\vspace{-.4cm}
\caption{\Bucketing. 
(1) Enclave registers a contract % {$P_1, P_2, P_3, P_4$}
(2) CPU directly reports the fault  to the \vdf. 
(3) \Vdf fakes access for time \texttt{t - k} and sends command to terminate.
(4) Enclave fault handler terminates the enclave.}
 % \vspace{-.5cm}
\label{fig:bucketing}
\end{figure} 

\paragraph{Mitigating Contract Violation.}
When the CPU signals contract violation and the control returns to the
enclave, it is important to terminate the program safely,  without leaking any
information (See Figure~\ref{fig:bucketing}).  When the enclave is notified
about contract violation, it is the enclaves responsibility to decide whether
to handle the fault or ignore it. One straightforward way to handle the fault
is terminate the enclave, but  our observation is that immediate program
termination leaks information (See Appendix~\ref{appxa}). In our solution, our
goal is to hide the following facts (a) whether the enclave incurred a page
fault during the execution after the contract is enforced (b) if so, at which
point  in the execution tree did the fault occur. To this end, in our defense
we intercept the page faults from the underlying hardware and from that point
of contract violation, we perform a fake execution to suppress the location at
which the fault happened. This defense can only work if we can ensure that the
enclave page fault handler is necessarily invoked. In the present SGX design
it is unclear if the hardware can guarantee the invocation of the page fault
handler. So we  propose that SGX can adopt this solution in the future. The
details of this mechanism are a bit involved and we for brevity we discuss it
in Appendix~\ref{appxb} for interested readers. We have implemented this
defense in \podarch and our evaluation on \libgcrypt~\footnote{We did not implement contractual execution for 
OpenSSL because it requires dynamic loading which is not supported in \podarch.} shows
that such an approach incurs an overhead of $6.77$\% which is much lower as
compared to the purely software based solutions (Table~\ref{tab:sec-perf}). We
elide the details here due to space limits. Please refer to Appendix~\ref{appxa} for details.

\subsection{Discussion: Other Alternative Approaches}
\label{sec:insuff}

\paragraph{Randomization of Page Access.} 
Oblivious RAM (ORAM) is a generic defense that randomizes the data access
patterns~\cite{goldreich1996oram, pathoram}.  Intuition suggests that the
enclave can use ORAM techniques to conceal its memory access pattern. In this
case, when an adversary observes the physical storage locations accessed, the
ORAM algorithm will ensure that the adversary has negligible probability of
learning anything about the true (logical) access pattern.  For our AES
example, we can place the tables in an ORAM to randomize their ordering, such
that the adversary cannot distinguish which offsets in the tables are
accessed.  However, ORAM involves continuous shuffling and re-encryption of
the data after every access. In our case studies, the lookup operations
dominate the computation in cryptographic implementations. For millions of
accesses, the cost incurred for the shuffling is significant poly log (say
over $1000\times$) and slows down the applications, which is not
desirable~\cite{shi2011oblivious}. Further, the best known ORAM technique
requires a constant private storage for shuffling the data blocks~\cite{const-oram}.
In case of \attack in SGX, the private storage is not permanently available to
the enclave and the OS can probe operations on private memory via page faults.
Thus, additional hardware support is necessary for ORAM based randomization to
justify the assumption of a secure constant private storage.

\paragraph{Self-Paging.} 
Instead of relying on the OS for page management, the enclaved execution can
take the responsibility of managing its memory. Applications can  implement
self-paging to deal with their own memory faults using their own physical memory
to store page tables~\cite{self-paging}.   In self-paging CPU design, all the
paging operations are removed from the kernel; instead the kernel is simply
responsible for dispatching fault notifications. Given a fixed amount of
physical memory, the enclave can decide which virtual addresses are mapped to
this memory, and which are swapped out. The problem with self-paging is ---
how can the enclave ensure that the  OS has allocated physical pages to it? To
guarantee this, the enclave should be able to pin certain physical memory
pages, such that the OS cannot swap them out. This directly opens the
possibility for a denial-of-service attack from the enclave, because it can
refuse to give up the pinned pages.  A hardware reset would be the only
alternative to reclaim all the enclave pages,  which is an undesirable
consequence for the OS.  Another possibility is that the enclave performs
self-paging without assuming fixed private physical memory. But this is
unsafe, since the OS still controls how much memory to allocate to the
enclave, retaining the ability to pigeonhole the memory pages. In both the
above alternatives, there is a dilemma --- should the enclave trust the OS and
likewise. Hence, it is unclear how self-paging, with or without fixed physical
memory, can defend against \attacks.
\section{Related Work} 
\label{sec:related} 

\paragraph{Attacks on Enclaved Execution.}
Xu \etal have recently shown that the OS can use the page fault channel on
applications running on SGX based systems to extract extract sensitive
information~\cite{cca-sgx}.  The attacks are limited to general user programs
such as image and text processing. On the contrary we study a cryptographic
implementations which is specific class of applications more relevant in the
context of enclaves. More importantly, we show that the purported techniques
discussed are not effective against \attacks. As as a new contribution, we
propose and measure the effectiveness of concrete solutions to prevent against
such attacks on cryptographic implementations. 

\paragraph{Side-channel Attacks.}
Yarom \etal study cache channel attacks wherein the adversary has the power to
flush and reload the cache, which can be used to attacks elliptic curve
cryptographic routines such as ECDSA~\cite{flush-reload, ecdsa-yarom}. Recent
study on caches has shown that even the last-level cache is vulnerable to
side-channel attacks~\cite{last-level-cache}. Timing and cache attacks have
been used to by-pass kernel space ASLR~\cite{kernel-aslr}, VMs~\cite{crossvm-aes}, 
android applications~\cite{memento}, cloud servers~\cite{paas-attack}
and users~\cite {hey-cloud} both locally and
remotely~\cite{Brumley03remotetiming}. Even web browsers can be exploited
remotely via cache attacks on JavaScript~\cite{spy-sandbox}.
 
\paragraph{Side-channel Detection \& Defenses.}
Various detection mechanisms have been explored for side channels ranging from
instruction level analysis to compiler techniques~\cite{ile-sidechannel,
cacheaudit, dangfeng-lang}. Tools such as CacheQuant can automatically
quantify the bits of information leaked via cache side-channels~\cite{kopf-cache-quant}. 
Techniques such as input blinding, time bucketing are also
available but are limited to specific algorithms~\cite{kopf-time-attack, kopf-smith}. 
Side channel attacks in hypervisors, cloud VMs, kernel are mitigated
using determinising strategies, control-flow independence and safe
scheduling~\cite{pc-security, deterministic-time, vm-channel-defense, duppel,stealthmem}. 
Our \interpreting defense is similar to memory-trace
obliviousness techniques proposed for secure computation~\cite{oblivm, csf13}.

\paragraph{Randomization \& Self-paging Defenses.}
ORAM techniques are widely used in secure computation and multi-party
computations. Recent work demonstrate safe language, compiler techniques, and
hypervisor based approaches which use ORAM.  As discussed in
Section~\ref{sec:insuff}, ORAM techniques may be insufficient without extra
hardware support.  On the other hand, self-paging assumes that the enclave
will always have control over a fixed size~\cite{self-paging}.  In case that
either party breaks this assumption, it opens a potential for DOS from enclave
and pigeonholing from the OS.
\section{Conclusion}
We systematically study \attack, a new threat prevalent in secure execution
platforms including Intel SGX, InkTag, OverShadow and  PodArch. By analyzing
cryptographic implementation libraries, we demonstrate the severity of
\attacks.  We  propose a purely software defense called \interpreting and
build a compiler to make all our case studies  safe  against \attacks. It is
practically deployable with modest overhead.   Finally, we present an
alternative hardware-based solution which incurs an average overhead of
$6.77$\%.

%ACKNOWLEDGMENTS are optional
\section{Acknowledgments}
This research is supported in part by the National Research Foundation, Prime Minister's Office, Singapore under its National Cybersecurity R\&D Program (Award No. NRF2014NCR-NCR001-21) and administered by the National  Cybersecurity R\&D Directorate.
%\end{document}  % This is where a 'short' article might terminate

% %ACKNOWLEDGMENTS are optional
% \section{Acknowledgments}
% This section is optional; it is a location for you
% to acknowledge grants, funding, editing assistance and
% what have you.  In the present case, for example, the
% authors would like to thank Gerald Murray of ACM for
% his help in codifying this \textit{Author's Guide}
% and the \textbf{.cls} and \textbf{.tex} files that it describes.

\bibliographystyle{abbrv}
\bibliography{paper}  % sigproc.bib is the name of the Bibliography in this case
% ACM needs 'a single self-contained file'!
% \balancecolumns
\appendix
%Appendix A
% \section{Safe \Bucketing}
\section{Safe Contractual Execution}
\label{appxa}

What should the enclave do once it detects that it is under attack or a
violation of the contract? 

\paragraph{Naive Self-termination Strategy}. 
A naive strategy is to immediately terminate the
\execution. Such deterministic self-termination by  the enclave leaks the
point of page fault to the OS, which leaks 1 bit of information per execution.
Note that the OS can repeatedly invoke the vulnerable application, stealing
different pages in each run and observing the different points of self-
termination in each execution. Such an adaptive pigeonholing adversary learns
significant information --- in fact, with enough trials the OS learns the same
amount of information by observing self-termination patterns as by observing
page faulting patterns in a vanilla enclaved execution!

For concrete illustration, consider the example of AES in which there are 4
input-dependent S-Box (table) lookup in each round. Let us consider the case
of two secret inputs  $I_1$ and $I_2$  such that execution under $I_1$ never
access a page $P_1$ in all of its S-box lookups, while  the execution under
$I_2$ accesses $P_1$ during the first S-box lookup.   To distinguish between
$I_1$ and $I_2$ in a contractual execution, the OS can steal  the data page
$P_1$ before the 3rd S-box lookup and observes whether the enclave self-
terminates abruptly. If it does, the OS can infer that the secret input is
$I_2$. Thus, abrupt termination serves as an oracle for the OS to distinguish
between two inputs $I_1$ and $I_2$. Specifically, the OS observe two things in
case of such termination: (a) the enclave was trying to access $P_1$, (b) the
index for 3rd lookup is less than \texttt{0x1c} since it accessed $P_1$.
Hence, deterministic self-termination is  not safe strategy.

To address the limitation, we introduce the notion of a working
set of pages. Each sensitive logic in the application defines a  minimum set
of physical pages that an enclave should have in the memory when executing it.
We refer to this working set as a {\em bucket}, whose size is specified in the
contract.  We first analyze the program execution tree and identify all the
code and data pages that  are accessed at all the levels of the execution
blocks. This defines the minimum required bucket size for a program.  At the
start of execution of a sensitive code area, the enclave initiates a contract
and requests the OS to commit to allocate  number of physical pages equal to
the size of bucket  (Step-1 in Figure~\ref{fig:bucketing} (a)). Once the bucket is
loaded in memory, the enclave executes the program assuming that the contract
is enforced.

In the event of a fault within a bucket, the CPU immediately vectors control
to the enclave's page fault handler.  It is the responsibility of the \vdf to
safely terminate the enclave. As mentioned earlier, the enclave cannot self-
terminate immediately when it detects that the bucketed page is missing. This
may reveal that the enclave was accessing the page and hence a particular code
branch / data in the path.

Once a contract violation is detected by the
enclave, the enclave enters into what we call as {\em fake execution} mode.
The fake execution mode is simply a spin-loop executed by the \vdf, which pads
the execution time of the program until it reaches the end of the execution
tree. In essence, the fake execution executes dummy blocks to mask the time of
occurrence of the page-fault from the OS. To execute this strategy, the \vdf
needs to know the time remaining (or elapsed) in the bucket execution. This
information is kept in a dedicated register during the program execution, and
is updated at the end of each block. The \vdf calculates the remaining time to
execute till the end of the tree using the information in the dedicated
register.  Figure~\ref{fig:bucketing} (a) shows the point at which the contract
violation occurs (Step 2), the fake execution (Step 3), and the termination (Step 4).

% The main challenge is that since the
% enclave cannot continue the real execution, what should the enclave do to make
% its ensuing execution  indistinguishable from a real execution? We now
% explain how to achieve such an indistinguishable fake execution. Before
% beginning a sensitive execution, the enclave initiates a contract and the OS
% loads all the bucket pages.  If the enclave detects an attack during the
% execution, the control enters \vdf. At this point, the \vdf knows the level of
% execution block within which the OS injected a fault (via the faulting
% address and execution  context information if necessary).  
% %This last step is crucial to achieve indistinguishably as we discuss below.

For such a defense to be secure, the execution of the enclave must be
indistinguishable to the adversary, independent of its strategy to respect or
violate the contract. Our described strategy achieves this goal.  Consider
three scenarios: (a) the OS obeys the contract, (b) the OS deviates from the
contract resulting in one or more page faults, and (c) the OS deviates from
the contract but no page faults result. Our defense ensures that all such
three executions are indistinguishable from the adversary's perspective. The
enclaves performs a real execution in scenario (a) and (c) and a fake
execution only in (b). All real executions incurs no page fault in contractual
execution --- hence they are indistinguishable trivially. It remains to show
that the fake execution is indistinguishable from a real execution.
Specifically, the time taken by the fake execution is the same as that by a
real execution (as explained above). Further, since all page faults are
redirected to the enclave, the OS does not see faults for the bucket pages,
and does not learn the sequence of the faulting addresses.  This establishes
that a fake execution is indistinguishable from the set of real executions.

\section{Preventing Pigeonhole attack ON enclave page fault handler}
\label{appxb}
To ensure that the \vdf can execute our strategy outlined above, the hardware
must guarantee a mechanism to vector control to a \vdf.  The current SGX
specifications have a mechanism to notify the enclave when  there is a page
fault, so that the enclave can implement its own page fault handler. However,
it does not specify whether the hardware guarantees that the  enclave's page
fault handler code will be mapped in memory when the enclave is executing.  If
this guarantee is missing, then a fault on the \vdf will lead to a double-
fault --- the accessed page as well as the fault handler page are
missing~\cite{podarch}.  To mitigate this threat, the hardware must eliminate
double-faults by design. Informing the OS about a double fault in unsafe, as
it leaks the information that \vdf was invoked thereby making fake execution
clearly distinguishable.

To prevent this leakage, we propose that the CPU allow  the enclave to specify
one virtual page in its contract  to always be mapped during its execution.
Specifically, the CPU checks if that this page is mapped whenever control
enter the enclave (say in the start of enclave execution or subsequently after
a context-switch).  Note this our proposal for ``pinning'' a page is different
from self-paging --- in our defense, the OS is free to invalidate the contract
by taking away the reserved page. This will result in the enclave being
aborted as soon as the context switches to the enclave, whether or not the
enclave accesses the reserved page. This abort strategy is thus independent of
page accesses in the enclave, and at the same time, the enclave  poses no risk
of denial-of-service to the OS.  We recommend this as an extension to enclave
systems such as SGX. 

\section{Details of our Attacks}
\label{appx:attacks}

% \subsection{Input Dependent Code Page Access}

\paragraph{\eddsa.}
In Section~\ref{sec:example} we explained how the page fault pattern for
scalar multiplication in \eddsa leaks value of $r$ completely. To use \eddsa,
the two parties first agree upon the public curve (or domain) parameters to be
used. For message $M$, the signing  algorithm outputs a tuple $(R,S)$ as the
signature. Specifically, the sender derives a session key $r$ for $M$ and uses
with a private key $a$. Here, the value of $r$ is used in the scalar
multiplication  operation $r \times G$, where $G$ is the public elliptic curve
point. In the verification step, the receiver checks if $\langle M, (R, S)
\rangle$ is a valid signature. If the adversary knows the value of $r$, he can
recover the value of $a$, and  easily forge signatures for any message $M$.

% The adversary can use the extracted  session key
% $r$ to retrieve the long term key $a$ by reverse engineering the relation $a =
% (S - r ) / (H(R, A, M)~mod~l)$,  where, A, H and l are public and $(R,S)$ is
% the known signature for M.  For an arbitrary message $M'$, the attacker uses
% private key $a$, picks a random session  key $r'$ and computes $R' = r'G$, $S'
% = r' + H(R', A, M')~a~(mod~l)$ to forge a new signature $(R', S')$.

\paragraph{\texttt{\bf powm}.}
Modular exponentiation (also referred to as \powm) is a basic operation to
calculate $g^d \bmod p$. It is used in many public-key cryptographic routines
(for e.g., RSA, DSA, ElGamal). Specially during key generation, decryption and
signing it involves a secret exponent (private key). Algorithm~\ref{alg:powm}
shows the outline of \powm implementation in \libgcrypt v1.6.3. \powm uses a
sliding window technique for exponentiation. This is essentially a m-ary
exponentiation which partitions the bits of the exponent into constant length
{\em windows}. The algorithm then performs as many multiplications as there
are non-zero words by sliding the window across the trailing zeros. The actual
\powm function body, the multiplication function and the selection function
are located in three separate pages. By the virtue of this, the OS can clearly
identify each call to a multiplication and a selection using the \profile.

Let us see the case when the window size $W= 1$. As the readers will observe,
there are two multiplication operations, one in the inner loop and one in the
outer loop, as highlighted in  Algorithm~\ref{alg:powm}. In order to know the
exact value of the secret exponent, it is important to identify in which loop
a particular multiplication operation is invoked. If the adversary can
distinguish each time the multiplication is invoked in the inner loop, it
effectively tells the number of 0 bits that the execution shifted after a bit
$1$.  To differentiate the inner loop multiplication ($a \cdot a$) from the
outer loop multiplication ($a \cdot g_u$), the adversary uses the following
strategy. It observes when $g_u$ is fetched from the precomputed table. To do
this, the algorithm invokes a bit checking logic and matches the value of $u$
with the corresponding value from the list of precomputed values. Since the
logic for  \texttt{set\_cond} is located in a different page, by observing the
page sequence, the attacker can group each individual multiplication to the
loop it belongs to. This leaks all the bits ($1$ separated by string of $0$s)
in the exponent in a single execution. Figure~\ref{fig:powm-example} (a) shows
the exact page fault pattern for \powm. The adversary needs approximately $W ^
{R}$ trials to extract the whole key, where  $R$ is the number of iterations
in the outer loop.

\paragraph{AES.}
As discussed in Section~\ref{sec:example}, there are $4$ T-Box tables in total
each with $256$ values, of which $2$ are split across pages., For the first
block of plain text, the routine directly uses the first $128$ bits of cipher
key in first round~\footnote{We trun off the Intel AES-NI hardware
acceleration in this case.}. The initial uncertainty of the OS is $2^{128}$.
With \attack, the OS knows for $64$ bits if the index is less than
\texttt{0x1c}  because they are for lookups in vulnerable S-Boxes. Thus, the
OS only needs to make  $2^{64} \times 28^{8}$ guesses. Thus, the information
leakage (in bits) = ${log_2}$(Initial Uncertainty -- Remaining Uncertainty).
${log_2}({2^{128}} - (2^{64} \times 28^{8})) = 25.54$ \textasciitilde{} $25$
bits~\cite{qif-smith}.  Thus, AES leaks $25$ bits in first rounds for all key
sizes $128$, $192$, $256$. 

\paragraph{Others.}
We use similar calculations steps for  the remaining cases --- \cast,
\seed, Stribog, Tiger and Whirlpool.
\seed  leaks $22$ bits and \cast leaks
$2$ bits in both \libgcrypt and \openssl. % (Table~\ref{tab:attacks}). For
\libgcrypt, the cryptographic hash implementations Whirlpool, Stribog and
Tiger leak $32$, $32$ and $4$ bits of the input key respectively in Password-
Based Key Derivation Function \texttt{PBKDF2} (Table~\ref{tab:attacks}).

% Note that the salt and randomization used in PKBDF2 is insufficient to prevent the
% attack because the vulnerable page fault patterns are exhibited even before
% they are used in the program logic~\cite{rfc2898}. 

% \paragraph{Hash Function Implementations.}
% PBKDF2 (Password-Based Key Derivation Function 2) is a key derivation function
% that takes in as input a  pseudo random function (\texttt{PRF}), the master
% password from which a derived key is to be generated (\texttt{password}), a
% cryptographic salt (\texttt{S}), the number of iterations desired (\texttt{c})
% and the desired length of the derived key (\texttt{dkLen}) such that:\\
% \texttt{DK = PBKDF2(PRF, Password, Salt, c, dkLen)} \\
% where, \texttt{DK} is the generated derived key.
% \libgcrypt implementation of PBKDF2 password expansion algorithm allows the
% user to configure the PRF from the available list of hash algorithms. $3$ of
% these implementations are vulnerable to \attacks because of lookup tables.
% This leaks the secret password used as input for the key expansion step 
% for the following hash implementations. 
\balancecolumns % GM June 2007
% That's all folks!
\end{document}